\documentclass[
lengthcheck,
amsmath,amssymb,
aps,
prl,
floatfix,
nofootinbib,
twocolumn,
]{revtex4-2}

\usepackage{graphicx}
\usepackage{lipsum}
\usepackage{dcolumn}
\usepackage{bm, bbm}
\usepackage[colorlinks, allcolors=blue]{hyperref} 
\def\equationautorefname#1#2\null{Eq.#1(#2\null)}

\usepackage{microtype}
\usepackage{parskip}
\usepackage{float}
\usepackage[dvipsnames]{xcolor}
\usepackage[caption=false]{subfig}
\begin{document}

\title{There is no ultrastrong coupling with photons}

\author{Diego Fernández de la Pradilla}
    \email{diego.fernandez@uam.es}
    \affiliation{Departamento de Física Teórica de la Materia Condensada and Condensed Matter Physics Center (IFIMAC), Universidad Autónoma de Madrid, E-28049 Madrid, Spain}
\author{Esteban Moreno}%
    \email{esteban.moreno@uam.es}
    \affiliation{Departamento de Física Teórica de la Materia Condensada and Condensed Matter Physics Center (IFIMAC), Universidad Autónoma de Madrid, E-28049 Madrid, Spain}
\author{Johannes Feist}%
    \email{johannes.feist@uam.es}
    \affiliation{Departamento de Física Teórica de la Materia Condensada and Condensed Matter Physics Center (IFIMAC), Universidad Autónoma de Madrid, E-28049 Madrid, Spain}
\date{\today}

\begin{abstract}
    Theoretical accounts of ultrastrongly coupled light-matter systems commonly assume that it arises from the interaction of an emitter with propagating photon modes supported by a structure, understanding photons as the excitations of the transverse electromagnetic field. This description discards the Coulomb interaction between the emitter and structure charges. Here, we show with a general argument based on electromagnetic constraints that the emitter-photon coupling strength is fundamentally limited. Accordingly, we conclude that the ultrastrong coupling regime cannot be reached with photons. Instead, it must originate from the Coulomb interactions between charges. A further corollary is that the so-called polarization self-energy term does not need to be included. We illustrate our claims by solving an analytical model of the paradigmatic case of an emitter next to a metallic nanosphere. These findings shed light on the fundamental processes underlying ultrastrong coupling, clarify the role of the polarization self-energy term and compel a reevaluation of previous literature.
\end{abstract}

\maketitle

\allowdisplaybreaks

The smallness of the fine structure constant, \(\alpha\simeq\frac{1}{137}\), implies that the interaction between quantum emitters and the electromagnetic (EM) field in free space is relatively weak. 
It has long been recognized that this limitation can be overcome by modifying the EM environment to reshape and enhance the field~\cite{purcell1946}.
In the extreme limit that the interaction strength \(\hbar g\) between a quantum emitter and an EM mode approaches the excitation energy of the emitter, \(\hbar\omega_\mathrm{e}\), the system enters the so-called ultrastrong coupling (USC) regime~\cite{kockumUSClightMatter2019,forndiazUSC2019}, conventionally defined by \(\frac{g}{\omega_\mathrm{e}} \geq 10^{-1}\).
The accompanying hybridization of states with different excitation numbers has various fundamental consequences, such as dressing of the ground state by a cloud of virtual photons~\cite{ciutiUSC2005}, leading to quantum-vacuum radiation when the system's parameters are rapidly modulated~\cite{günterSwitchOnUSC2009,garzianoSwitchOnOff2013}.
USC has also been proposed for use in preparing non-classical states~\cite{ashhabNonClassical2010} or modifying ground-state material and chemical properties~\cite{schwartzReversibleUSC2011, wangPhaseTransition2014, Garcia-Vidal2021, Lu2024}.
Additionally, the USC regime distinctly affects the photon emission statistics from the system~\cite{deliberatoVacuumRadiation2007, ridolfoPhotonBlockade2012, ridolfoNonclassicalUSC2013, cirioElectrolum2016}.
Due to these effects, USC physics has the potential to become relevant for technological applications involving chemistry~\cite{martinezUSCchemistry2018}, materials science~\cite{scalariUSC2DEG2012}, and quantum information~\cite{romeroQuantumGate2012, felicettiPhotonTransfer2014, kyawErrorCorrectionUSC2015, xiongUSCplexcitonic2020}.
Crucially, most predicted effects depend on the per-emitter coupling strength and do not benefit from collective enhancement---in other words, they require a \emph{single} emitter to couple ultrastrongly to a cavity mode.
In the following, we thus focus on the single-emitter USC regime.

\begin{figure}[t]
    \centering
    \includegraphics[width=0.8\linewidth]{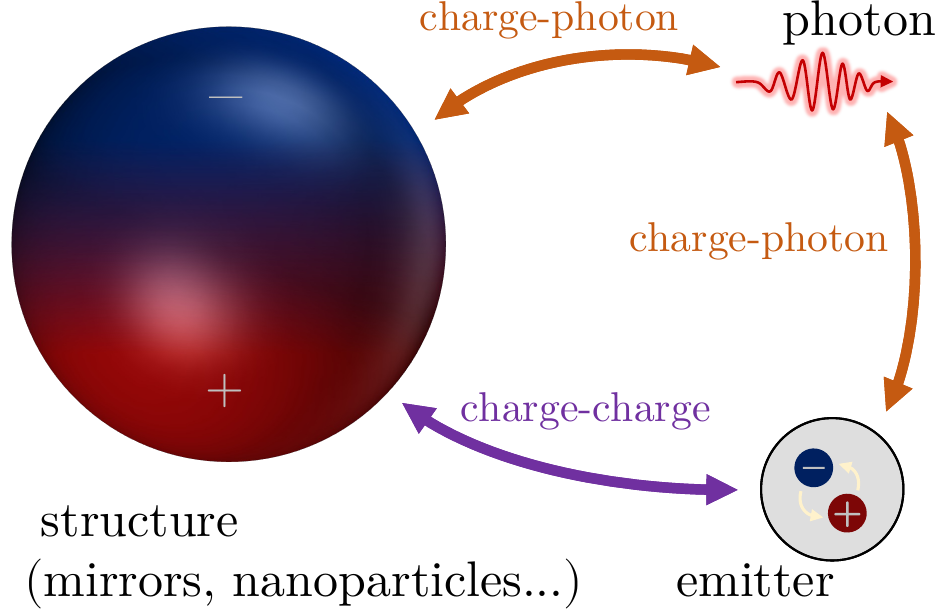}
    \caption{
        Sketch of a general QED setup with two sets of charges (emitter and nanophotonic structure) coupled to photons, with all three parts interacting with each other.
        }
    \label{fig.scheme_intro}
\end{figure}

Much of the literature on USC assumes that it arises from the coupling of an emitter to a \emph{photonic} cavity mode.
With ``photon'', we refer to excitations of the true dynamical degrees of freedom of the EM field, described by the transverse (divergence-free) components \(\mathbf{E}^\perp(\mathbf{r})\) and \(\mathbf{B}^\perp(\mathbf{r})\). 
Using the appropriate complete theory, non-relativistic QED~\cite{craigMolecularQED1984,cohenQEDIntro2007}, we show here that this assumption is fundamentally flawed, and that in fact USC cannot be reached by coupling between a single emitter and photons alone.
This follows from a sum rule capturing the fact that the maximum density of states of photons is limited since the presence of a cavity can only rearrange these states, but not create new ones.
Instead, single-emitter USC can only be reached through the Coulomb interaction between an emitter and the charges within the material structure that hosts the cavity mode (schematically depicted in \autoref{fig.scheme_intro}). We show that the contribution of photons is negligible in all cases.
In addition to the general argument, we illustrate our findings through an analytically solvable example where the separation between charge-charge and charge-photon interactions is transparent.

The above statements have important implications for the description of USC\@. 
The common but flawed assumption of purely photonic cavity modes implies that the form of the coupling depends on the choice of gauge~\cite{cohenQEDIntro2007}, and in particular that the light-matter coupling includes a term proportional to the square of either the EM vector potential, \(\propto g^2 \mathbf{A}^2\), or of the polarization density of the emitter, \(\propto g^2 \mathbf{P}^2\) (which in the long-wavelength approximation turns into the so-called dipole self-energy, \(\propto g^2 \mathbf{d}^2\), with \(\mathbf{d}\) the emitter dipole operator).
The presence of these terms can have important consequences, such as an effective decoupling of light and matter for large enough coupling~\cite{DeLiberato2014, Garcia-Ripoll2015} or significant modifications of the chemical structure of molecules~\cite{rokajNoGroundState2018, schäferAsquared2020, Borges2024, weightBromination2024, schnappingerMolecularGeometries2024}.
In contrast, the charge-charge (Coulomb) interactions between the cavity structure and the emitter that we show to be responsible for USC can be described through the longitudinal (curl-free) component of the electric field, \(\mathbf{E}^\parallel(\mathbf{r})\), which is not a dynamical degree of freedom but rather fully determined by the instantaneous position of the charges.
This interaction does not give rise to \(\mathbf{A}^2\) nor \(\mathbf{P}^2\)-terms, such that these cannot play an important role in physical setups.

We note the that statements above are not necessarily surprising.
For example, it is well-known that subwavelength field confinement is required to reach single-emitter USC~\cite{benzSingleMolecule2016, grossNearStrong2018, kuismaUSCsingleMoleculePlasmonic2022}, and that subwavelength field confinement implies that the longitudinal fields dominate over the transverse ones~\cite{Khurgin2015, galegoCPforces2019}.
Still, we are not aware of any work that has explicitly demonstrated the bound on the emitter-photon coupling strength that we present here. In fact, much of the existing literature on USC overlooks this bound by assuming that USC is reached through the emitter-photon interaction. We thus believe that our findings are important to clarify the fundamental processes underlying USC, and to compel a reevaluation of previous literature.

\section*{Theoretical framework} \label{sec.argument}

In this section, we frame the problem of USC in terms of charges and photons within non-relativistic QED~\cite{cohenQEDIntro2007,craigMolecularQED1984}. 
We next apply this description to an emitter coupled to the EM modes supported by an arbitrary material structure, which we will call the ``cavity'' for convenience regardless of its physical nature (which can be a photonic crystal, a plasmonic structure, etc.).
Then, we show that USC can only be reached when the dominant mechanism is the charge-charge interaction between the emitter and cavity.
We start from the minimal coupling Hamiltonian in the Coulomb gauge (\(\nabla\cdot\mathbf{A}=0\)), which describes the charges of both the emitter and the cavity material, and the EM field~\cite{cohenQEDIntro2007},
\begin{align} \label{eq.H_general_minimal}
    H &= H_\mathrm{m} + H_\mathrm{f} + H_\mathrm{int}\\
    \nonumber
    H_\mathrm{m} &= \sum_i \frac{\mathbf{p}_i^2}{2m_i} + \sum_{i>j} \frac{q_i q_j}{4\pi\varepsilon_0|\mathbf{r}_i - \mathbf{r}_j|} \\
    \nonumber
    H_\mathrm{f} &= \int\mathrm{d}^3r\ \left(\frac{\left(\varepsilon_0\mathbf{E}^\perp(\mathbf{r})\right)^2}{2\varepsilon_0} + \frac{\varepsilon_0 c^2}{2}\left(\nabla\times\mathbf{A}^\perp(\mathbf{r})\right)^2\right) \\
    \nonumber
    H_\mathrm{int} &= - \sum_i \frac{q_i }{m_i}\mathbf{p}_i\cdot \mathbf{A}^\perp(\mathbf{r}_i) + \sum_i \frac{q_i^2}{2m_i} \left(\mathbf{A}^\perp(\mathbf{r}_i)\right)^2.
\end{align}
Here, \(H_\mathrm{m}\) is the matter Hamiltonian for all point charges (electrons and nuclei), characterized by their masses \(m_i\), charges \(q_i\), positions \(\mathbf{r}_i\), and momenta \(\mathbf{p}_i\), which interact with each other through the Coulomb potential.
The Hamiltonian \(H_\mathrm{f}\) describes the dynamical EM field components, which are represented through the Coulomb vector potential \(\mathbf{A}^\perp\) and its corresponding canonical momentum \(-\varepsilon_0\mathbf{E}^\perp\). Because we have chosen the Coulomb gauge, the vector potential is equal to its gauge-invariant transverse component, as denoted with the superscript. 
\(H_\mathrm{int}\) in the last line mediates charge-photon interactions. 
We remark that the separation between longitudinal and transverse electric fields (as opposed to potentials) does not depend on gauge choices. In that sense, the Coulomb gauge is the one in which the potentials most closely mirror this physics.
We also note that the Hamiltonian is invariant under translations of all the charges, as one would expect.
However, it is not invariant under translations of just the emitter while keeping the cavity fixed in space. Requiring this invariance, as sometimes done in abstract treatments based on single-mode Hamiltonians~\cite{rokajNoGroundState2018, schäferAsquared2020}, is thus not physically meaningful.

For emitters in free space, the light-matter coupling is weak enough that the EM environment simply induces small energy shifts (Lamb shift) and decay rates on the emitter levels (after proper renormalization of diverging integrals due to the point-like nature of the charges)~\cite{buhmannDispersionI2012}.
Even so, since the charges and the EM field are coupled, the eigenstates of the Hamiltonian are at least formally hybrid polaritonic states that describe the correlated motion of charges and photons. A cavity is then, from this viewpoint, simply a large enough assembly of charges that supports approximately bosonic ``cavity modes'' that provide the new effective EM environment for the emitter.
The use of bosonic modes implies that the cavity material is treated within linear response and thus effectively corresponds to a collection of harmonic oscillators that hybridizes with the free-space EM modes to form the polaritonic cavity modes.
We note that within this framework, even the modes of Fabry-Pérot microcavities, i.e., standing waves between two mirrors, are properly understood as polaritons, as the reflection of the EM fields by the mirrors is due to the reaction of the charges to the fields.
Naturally, these hybrid modes generate both longitudinal fields, \(\mathbf{E}^\parallel\), and transverse ones, \(\mathbf{E}^\perp\).

Focusing on the emitter, its coupling to the rest of the system can be written as
\begin{align}\label{eq.H_int_emitter_medium_assisted}
    H_\mathrm{int,e} = -\int\mathrm{d}^3r\ \mathbf{P}_\mathrm{e}(\mathbf{r})\cdot\mathbf{E}_\mathrm{c}^\parallel(\mathbf{r}) - \sum_{i \in \mathrm{e}} \frac{q_i }{m_i}\mathbf{p}_i\cdot \mathbf{A}^\perp(\mathbf{r}_i),
\end{align}
which shows an unambiguous separation between longitudinal and transverse interactions. The first, longitudinal, term is the Coulomb interaction with the cavity charges rewritten (through integration by parts) in terms of the emitter polarization density, defined by \(\nabla\cdot\mathbf{P}_\mathrm{e}(\mathbf{r}) = -\sum_{i\in \mathrm{e}}q_i\delta(\mathbf{r}-\mathbf{r}_i)\), and the longitudinal electric field \(\mathbf{E}_\mathrm{c}^\parallel(\mathbf{r}) = \sum_{i\in \mathrm{c}} \frac{q_i}{4\pi\varepsilon_0} \frac{\mathbf{r} - \mathbf{r}_i}{|\mathbf{r} - \mathbf{r}_i|^3}\) generated by the cavity charges.
The second term is the interaction with the transverse fields (i.e., photons), which depends on \(\mathbf{A}^\perp\).
Although it is formally identical to its free-space counterpart in \autoref{eq.H_general_minimal}, the cavity structure implicitly modifies the transverse interaction as well, because it alters the propagation properties of photons and thus affects the evolution of \(\mathbf{A}^\perp\).

We now focus on the transverse interaction in \autoref{eq.H_int_emitter_medium_assisted} and manipulate it to obtain more explicit expressions.
To that end, we expand \(\mathbf{A}^\perp\) in free-space photonic modes:
\begin{equation}\label{eq.A_expansion}
    \mathbf{A}^\perp(\mathbf{r}) = \sum_\lambda A_\lambda \mathbf{f}_\lambda(\mathbf{r}).
\end{equation} 
The mode functions \(\mathbf{f}_\lambda\) are solutions of the Helmholtz equation with frequency \(\omega_\lambda\), are normalized to \(\int\mathrm{d}^3r\ \mathbf{f}_\lambda(\mathbf{r}) \cdot \mathbf{f}_{\lambda'}(\mathbf{r}) = \delta_{\lambda\lambda'}\), and have been chosen real without loss of generality.
They form a complete orthogonal basis for the space of transverse vector fields.
Each basis function is accompanied by the corresponding free-space photon mode displacement operator \(A_\lambda = \sqrt{\frac{\hbar}{2\varepsilon_0\omega_\lambda}}\left(a_\lambda + a_\lambda^\dagger\right)\), where \(a^{(\dagger)}_\lambda\) is the corresponding annihilation (creation) operator.
Although the notation of a single sum over a combined index \(\lambda\) suggests a countable number of modes, this is just chosen for simplicity and generality of notation---any specific choice of basis in (infinite) free space will involve at least one continuous index for which the sum becomes an integral and the Kronecker delta becomes a Dirac delta function.
The emitter-photon interaction Hamiltonian is then
\begin{align}
    \nonumber
    H_\mathrm{int,e}^{\perp} &= - \sum_{i\in\mathrm{e}} \frac{q_i}{m_i}\mathbf{p}_i\cdot\sum_\lambda A_\lambda \mathbf{f}_\lambda(\mathbf{r}_i)\\
    \label{eq.H_int_transverse_photon}
    &\simeq i\hbar\sum_t \sigma_{t} \sum_\lambda \frac{\omega_t}{\omega_\lambda} g^{\perp}_{t\lambda,0} \left(a_\lambda^\dagger + a_\lambda\right) + \mathrm{H.c.},
\end{align}
In the second equality of \autoref{eq.H_int_transverse_photon}, we expanded the emitter operator in its transitions \(t\) with frequency \(\omega_t\geq0\) and lowering operators \(\sigma_t\).
Additionally, we use that \(\mathrm{Tr}\{\sigma_t^\dagger\mathbf{p}_i\} = -i\omega_t m_i \mathrm{Tr}\{\sigma_t^\dagger\mathbf{r}_i\}\) to retrieve the transition dipole moment \(\mathbf{d}_t = \sum_{i\in\mathrm{e}} q_i \mathrm{Tr}\{\sigma_t^\dagger\mathbf{r}_i\}\) and define the photonic transverse coupling strength
\begin{equation}
    \label{eq.transverse_photon_coupling}
    g_{t\lambda,0}^{\perp} = \sqrt{\frac{\omega_\lambda}{2\hbar\varepsilon_0}} \mathbf{d}_t\cdot \mathbf{f}_\lambda(\mathbf{r}_\mathrm{e}),
\end{equation}
where \(\mathbf{r}_\mathrm{e}\) denotes the emitter's position. 
For simplicity, we used the long-wavelength approximation, though it is not essential for the argument's validity.
Note that \(g_{t\lambda,0}^{\perp}\) corresponds to the ``bare'' coupling in the multipolar picture and appears naturally in the sum rules discussed below, but occurs with a factor \(\omega_t/\omega_\lambda\) in \autoref{eq.H_int_transverse_photon}.
Whether including the frequency factor or not predicts the (picture-independent) level splitting for non-resonant cases more accurately depends on the emitter structure~\cite{debernardisgaugebreakdown2018}.

The coupling of the cavity charges to the free-space photon modes modifies the dynamics of the photon displacement operators \(A_\lambda\).
Consequently, the clearest way to describe the emitter-photon and emitter-cavity interactions is to diagonalize the photon-cavity subsystem and write the Hamiltonian in terms of the new uncoupled polaritonic mode displacement operators \(\beta_\eta\) with frequency \(\omega_\eta\). 
Defining polaritonic creation and annihilation operators as \(\beta_\eta = \sqrt{\frac{\hbar}{2\omega_\eta}}\left(b_\eta + b_\eta^\dagger\right)\), the transverse interaction Hamiltonian can be rewritten as
\begin{equation}
    H_\mathrm{int,\mathrm{e}}^{\perp} = i\hbar\sum_t \sigma_{t} \sum_\eta \frac{\omega_t}{\omega_\eta} g^{\perp}_{t\eta} \left(b_\eta^\dagger + b_\eta\right) + \mathrm{H.c.},
\end{equation}
with the polaritonic transverse coupling strength
\begin{align}
    \label{eq.transverse_polariton_coupling}
    g_{t\eta}^{\perp} &= \sum_\lambda g_{t\lambda,0}^{\perp} M_{\lambda\eta}.
\end{align}
Here, \(M_{\lambda\eta}\) is the diagonalization matrix that relates \((b_\eta^\dagger + b_\eta)\) to \((a_\lambda^\dagger + a_\lambda)\) and projects the interaction onto the polaritonic modes. Explicitly finding \(M_{\lambda\eta}\) is usually not trivial, so we keep the discussion abstract here (see the last section and the methods for a concrete analytical example).
For our purposes, it is convenient to introduce the transverse spectral densities \(J_t^\perp\) and \(J_{t,0}^{\perp}\), an alternative way to characterize the coupling strength:
\begin{equation} \label{eq.spectral_def}
    J_{t(,0)}^{\perp}(\omega) = \sum_{\mu} \left(g_{t\mu(,0)}^{\perp}\right)^2 \delta(\omega - \omega_\mu).
\end{equation}

\section*{Derivation of the bound}
We next derive a bound on the emitter-photon coupling strength by leveraging certain constraints on how \(M\) can distribute the coupling strength across frequencies. 
As we will demonstrate, two such restrictions impose a quantitative limitation on the coupling due to the \(\mathbf{p}_i\cdot\mathbf{A}^\perp(\mathbf{r}_i)\) term. 
The first one is that any material becomes transparent at sufficiently high frequencies, that is, \(J_t^\perp(\omega) \to J_{t,0}^{\perp}(\omega)\) as \(\omega \to \infty\).
The second restriction arises from a sum rule due to the properties of \(M_{\lambda\eta}\), derived in the supplementary information. In terms of the spectral density, it can be expressed as
\begin{equation}
    \label{eq.sum_rule}
    \int_0^\infty\mathrm{d}\omega\ \frac{J_{t}^\perp(\omega) - J_{t,0}^{\perp}(\omega)}{J_{t,0}^{\perp}(\omega)} = 0,
\end{equation}
which means that the transverse relative coupling enhancement (equal to the Purcell factor) integrated over a sufficiently large frequency range averages to 1. We note that this sum rule, derived here in non-relativistic QED, can also be obtained for the classical density of states in macroscopic electromagnetism~\cite{barnettSumRule1996} and, within macroscopic QED~\cite{buhmannMQED2008}, directly extends to the quantum case.

To obtain a bound for the maximum coupling to transverse modes, let us imagine an idealized ``perfect'' transverse cavity, i.e., one that concentrates all photonic modes up to a ``transparency'' frequency \(\Omega_T\) into a single transverse cavity mode in resonance with an emitter transition \(t_0\). Beyond \(\Omega_T\), the free-space modes remain essentially uncoupled from the structure, in accordance with the first constraint. 
Hence, the spectral density is
\begin{subequations}
    \begin{equation}
        \label{eq.J_perfect_transverse_cavity}
        J_{t_0}^\perp(\omega) = \left(G^\perp_{t_0}\right)^2\delta(\omega-\omega_{t_0}) + J_{t_0, 0}^{\perp}(\omega) \theta(\omega-\Omega_T),
    \end{equation}
    where \(G^\perp_{t_0}\) is the coupling strength of the emitter transition \({t_0}\) to the transverse cavity mode, \(\theta\) is the step function and the free-space spectral density is
    \begin{equation}
        \label{eq.J_free_space}
        J_{t_0, 0}^{\perp}(\omega) = \frac{\left|\mathbf{d}_{t_0}\right|^2\omega^3}{6\pi^2\hbar\varepsilon_0 c^3}.
    \end{equation}
\end{subequations}
Introducing both in \autoref{eq.sum_rule} yields
\begin{equation}
    \label{eq.transverse_total_coupling}
    G^\perp_{t_0} = \left|\mathbf{d}_{t_0}\right|\sqrt{\Omega_T\frac{\omega_{t_0}^3}{6\pi^2\hbar\varepsilon_0 c^3}}.
\end{equation}

This expression is the main result of the article and implies that the coupling strength concentration in a single transverse cavity mode is limited by the value of the transparency frequency \(\Omega_T\).
To apply \autoref{eq.transverse_total_coupling} to the description of single- or few-emitter USC, we note that the total coupling strength also depends on the emitter's dipole moment.
Thus, we invoke the Thomas-Reiche-Kuhn sum rule to find that 
\begin{equation}
    \label{eq.dipole_trk}
    |\mathbf{d}_{t_0}| < \sqrt{\frac{3\hbar e^2}{2m_e\omega_{t_0}}n}.
\end{equation}
Here, \(e\) and \(m_e\) are the electron's charge and mass, and we can set \(n \sim 1\) because we are considering small emitters dominated by single- or few-electron physics. 
Combining \autoref{eq.transverse_total_coupling} and \autoref{eq.dipole_trk} yields
\begin{equation}\label{eq.G_perp_limit}
    \frac{G^\perp_{t_0}}{\omega_{t_0}} < \sqrt{\frac{\Omega_T e^2}{4 \pi^2 \varepsilon_0 m_e c^3} } \simeq \sqrt{\frac{\hbar\Omega_T}{220~\text{MeV}}},
\end{equation}
independent of the emitter frequency. We note that the energy scale of \(\simeq 220~\)MeV can be conveniently expressed in atomic units as \(\pi \alpha^{-3}~\)Hartree, demonstrating that the bound is indeed a consequence of the smallness of the fine-structure constant.
This result shows that USC (i.e., \(G^\perp_{t_0} / \omega_{t_0} \geq 0.1 \)) could only be reached if \(\hbar\Omega_T \geq 2~\)MeV, a value that lies several orders of magnitude above the frequencies at which real materials become transparent.
Put more explicitly, a hypothetical cavity that achieves USC through photon-emitter interactions would have to be able to perfectly reflect all photons with energies from zero up to larger than 2~MeV.
It cannot be overstated how unreasonably high this value is for any realistic material---broadband reflectivity is limited by the plasma frequency of free electrons in the material, which rarely exceeds 10~eV in known materials.
Furthermore, conventional cavity designs made with such a miraculous material would lead to cavity modes at frequencies on the order of MeV, and would require a picometer-sized cavity, smaller than a single atom.
It follows that single-emitter USC is well out of reach through photons alone.
Instead, the Coulomb interaction between charges in the emitter and cavity must be the physical mechanism that unlocks USC\@.

\section*{Role of the polarization self-energy}
We continue with a further corollary concerning the polarization self-energy. 
As mentioned in the introduction, the PSE is a term that appears in the multipolar coupling representation of the light-matter Hamiltonian, related to \autoref{eq.H_general_minimal} by the Power-Zienau-Woolley (PZW) transformation~\cite{powerPZW1959,woolleyPZW1971,cohenQEDIntro2007}. 
In the long-wavelength approximation, it is given by
\begin{equation}
    \int\mathrm{d}^3r\ \frac{\left(\mathbf{P}_\mathrm{e}^\perp(\mathbf{r})\right)^2}{2\varepsilon_0} \simeq \hbar\sum_\eta \frac{\left(\sum_t g_{t\eta}^{\perp}\sigma_t + \mathrm{H.c.}\right)^2}{\omega_\eta},
\end{equation}
where \(g_{t\eta}^{\perp}\) is the transverse emitter-mode coupling strength from \autoref{eq.transverse_polariton_coupling}. Note that even if the PSE can be expressed in terms of the emitter-mode couplings, the full sum over all the modes is cavity-independent, as it only depends on the emitter polarization \(\mathbf{P}^\perp_\mathrm{e}\).
Still, the PSE is often included in simplified models where the emitter couples to a single cavity mode \(a\) through its dipole operator \(\hat{d}\)~\cite{schäferAsquared2020,debernardisCQEDNonPert2018,distefanoResolutionGaugeAmbiguities2019,schnappingerCavityBornOppenheimer2023,schnappingerMolecularGeometries2024,weightBromination2024}: 
\begin{equation}
    H^\perp_\mathrm{simpl.} = H_\mathrm{e} + \hbar\omega a^\dagger a+\hbar G^\perp \hat{d} \left(a + a^\dagger\right) + 
    \hbar \frac{\left(G^\perp \hat{d}\right)^2}{\omega}
\end{equation}
This model, sometimes called the Pauli-Fierz Hamiltonian,
assumes that \(a\) is a purely photonic cavity mode.
When \(G^\perp\) is chosen large enough to have significant impact on the system, the single-mode PSE seems to induce a non-negligible renormalization of \(H_\mathrm{e}\).
However, the bound derived in \autoref{eq.G_perp_limit} limits the single-mode PSE as well, and effectively shows that it can never have an appreciable effect on the emitter (this is true even in the many-emitter case, since the PSE does not show a collective enhancement).
Instead, any significant mode-emitter coupling must be due to the longitudinal field \(\mathbf{E}^\parallel\) in \autoref{eq.H_int_emitter_medium_assisted}, expressed in terms of the polaritonic cavity mode operators. The corresponding simplified Hamiltonian is
\begin{equation}
    H_\mathrm{simpl.}^\parallel = H_\mathrm{e} + \hbar\omega a^\dagger a + \hbar G^\parallel \hat{d} \left(a + a^\dagger\right),
\end{equation}
without a PSE term. 
Previous works claimed that the PSE is crucial to ensure the existence of a bound state for sufficiently large computational boxes~\cite{schäferAsquared2020,rokajNoGroundState2018}, as the energy of a charge can become arbitrarily negative with increasing displacement.
In contrast, our discussion here indicates that few-emitter USC comes from charge-charge Coulomb interactions, where the PSE plays no role at all.
Since the Coulomb interaction is known not to prevent the existence of bound states, the non-existence of a bound state is revealed to be due to the additional approximations, in particular the long-wavelength approximation that effectively assigns a constant electric field in all of space to the ``cavity'' mode.
Then, a true physical fix requires going beyond the long-wavelength approximation and taking into account that the ``cavity'' field is not constant in space, or eventually including the subwavelength cavity in the ab-initio description.

\section*{Explicit analytic model}

\begin{figure*}[thbp]
    \centering
    \subfloat{\label{fig.system_sketch}}
    \subfloat{\label{fig.spectral_perp_long}}
    \subfloat{\label{fig.spectral_comparison}}
    \subfloat{\label{fig.longitudinality}}
    \subfloat{\label{fig.pse_potential}}
    \includegraphics[width=\textwidth]{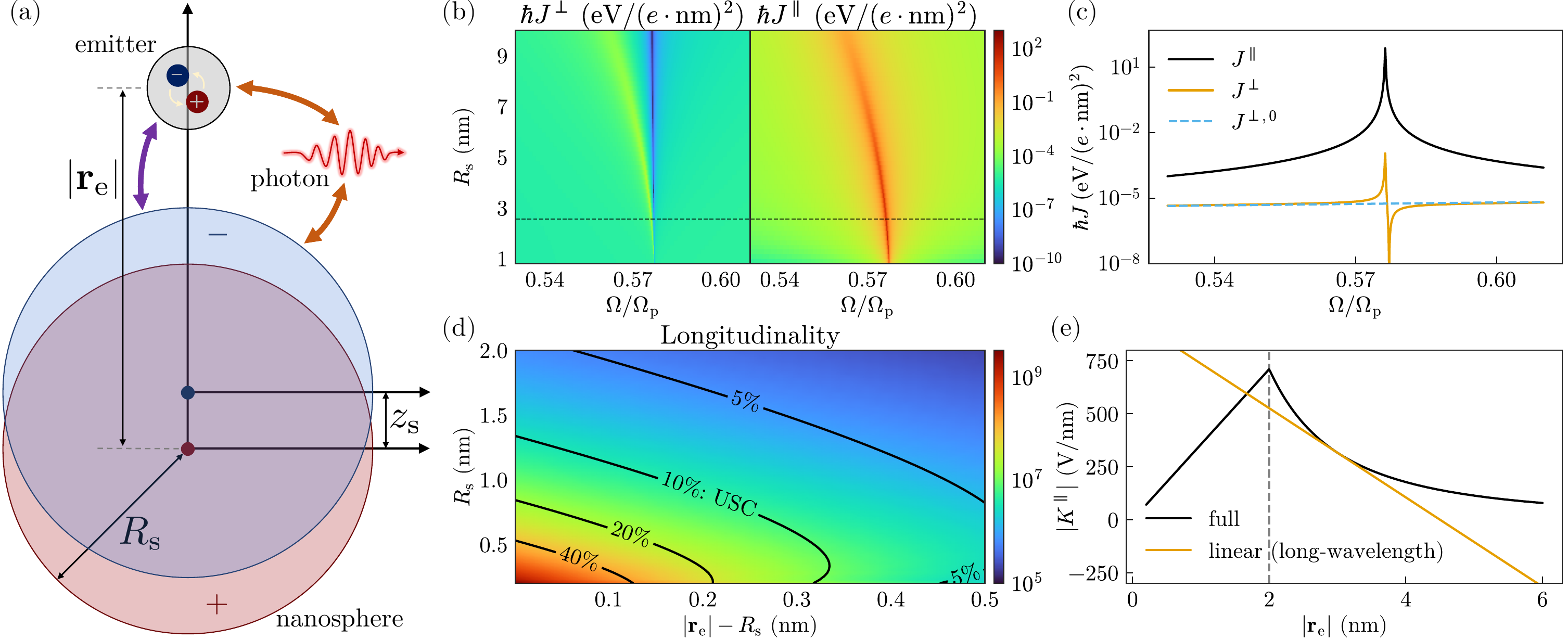} 
    \caption{
        (a) Sketch of the model. (b) Left: transverse spectral density \(J^\perp\). Right: longitudinal spectral density \(J^\parallel\). Both obtained per unit dipole moment squared, and as a function of the sphere's radius and the frequency, setting the emitter-nanoparticle separation to \(|\mathbf{r}_\mathrm{e}|-R_\mathrm{s}=1\)~nm. (c) Comparison of \(J^\parallel\) and \(J^\perp\) at \(R_\mathrm{s}\simeq2.5\)~nm, marked by the horizontal dashed line in (b). (d) Averaged ratio between the longitudinal and transverse spectral densities. The contour lines indicate the fraction of the excitation frequency reached by the coupling strength, for a transition dipole moment equal to 10 Debyes. (e) Full spatial dependence of \(\phi_\mathrm{s}\) and its long-wavelength approximation counterpart. The dashed vertical line marks \(R_\mathrm{s}\).
        }
    \label{fig.system_results}
\end{figure*}

We illustrate the conceptual discussion above with the paradigmatic setup of an emitter close to a plasmonic nanosphere acting as a cavity.
Here, we explain the most important modelling aspects and results, relegating the detailed calculations to the methods and supplementary information.
We represent the conduction electrons and background ions of the plasmonic particle as two overlapping spherical homogeneous charge distributions of radius \(R_\mathrm{s}\) with opposite charge.
The ionic sphere (charge density \(\rho\)) is fixed at the origin, while the electronic sphere (\(-\rho\)) moves along the \(z\) axis with displacement \(z_\mathrm{s}\). The Coulomb attraction (\autoref{fig.system_sketch}) then creates a harmonic potential for \(z_\mathrm{s}\ll R_\mathrm{s}\), with frequency \(\Omega_\mathrm{p}/\sqrt{3}\), where \(\Omega_\mathrm{p} = \sqrt{\frac{\rho e}{m_e \varepsilon_0}}\) is the plasma frequency. Note that \(\Omega_\mathrm{p}/\sqrt{3}\) is precisely the quasistatic dipolar resonance frequency of a small metallic sphere with a lossless Drude permittivity in vacuum, indicating that this is indeed a reasonable model for such a particle.
For typical values of \(R_\mathrm{s}=10\)~nm, \(\rho=60~e\cdot\mathrm{nm}^{-3}\) and total energies around 1~eV, \(z_\mathrm{s}\sim 1\)~pm and the small oscillation approximation thus holds. The Hamiltonian is
\begin{align} \label{eq.H_model}
    H &= H_\mathrm{e} + H_\mathrm{s} + H_\mathrm{f} + H_\mathrm{int, e}^\parallel \\
    \nonumber
    H_\mathrm{e} &= \sum_{i\in\mathrm{e}} \frac{\left(\mathbf{p}_i - q_i \mathbf{A}^\perp(\mathbf{r}_i)\right)^2}{2m_i} + \sum_{i>j\in\mathrm{e}} \frac{q_i q_j}{4\pi\varepsilon_0|\mathbf{r}_i - \mathbf{r}_j|} \\
    \nonumber 
    H_\mathrm{s} &= \frac{\left(p_\mathrm{s} + \displaystyle\int\mathrm{d}^3r\ \rho(\mathbf{r}) \hat{\mathbf{z}}\cdot \mathbf{A}^\perp(\mathbf{r})\right)^2}{2m_\mathrm{s}} + \frac{m_\mathrm{s}}{2}\frac{\Omega_\mathrm{p}^2}{3} z_\mathrm{s}^2 \\
    \nonumber
    H_\mathrm{f} &= \int\mathrm{d}^3r\ \left(\frac{\left(\varepsilon_0\mathbf{E}^\perp(\mathbf{r})\right)^2}{2\varepsilon_0} + \frac{\varepsilon_0 c^2}{2}\left(\nabla\times\mathbf{A}^\perp(\mathbf{r})\right)^2\right) \\
    \nonumber
    H^\parallel_\mathrm{int,e} &= - \int\mathrm{d}^3 r\ \mathbf{P}_\mathrm{e}(\mathbf{r})\cdot \mathbf{E}_{\mathrm{s}}^\parallel(\mathbf{r}) = \sum_{i\in\mathrm{e}} q_i \phi_\mathrm{s}(\mathbf{r}_i, z_\mathrm{s}),
\end{align}
where \(m_\mathrm{s}\), \(p_\mathrm{s}\) and \(-\rho(\mathbf{r}) = -\rho \theta(R_\mathrm{s} - |\mathbf{r}-z_\mathrm{s}\hat{\mathbf{z}}|)\), are the electronic sphere's mass, momentum and charge density, expressed with a step function, and \(\hat{\mathbf{z}}\) is the unit vector along the \(z\) axis.
The transverse interactions are included in the emitter and sphere Hamiltonians directly, where the integral in \(H_\mathrm{s}\) is a straightforward generalization of \(-\sum_i q_i\mathbf{A}^\perp(\mathbf{r}_i)\) for an extended charge distribution, and \(\phi_\mathrm{s}(\mathbf{r}, z_\mathrm{s})\) is the Coulomb potential due to the spheres (note that no long-wavelength approximation is made).

We now diagonalize the sphere-photon subsystem and find the emitter-eigenmode interaction, like in the conceptual discussion above.
After a somewhat lengthy procedure, detailed in the methods and supplementary information, we obtain a Hamiltonian that can be analytically diagonalized using the Fano technique~\cite{fanoDiagonalization1961}.
This allows us to express the transverse and longitudinal interactions with a two-level emitter, placed on the \(z\) axis with transition dipole moment along the same direction and frequency \(\omega_t\), in terms of the polaritonic eigenmodes.
The closed expressions for \(g^\perp(\omega)\) and \(g^\parallel(\omega)\) analogous to \autoref{eq.transverse_polariton_coupling} are given in the methods.
Here, we focus on the corresponding spectral densities, essentially \(\left(g^{\perp/\parallel}(\omega)\right)^2\), plotted in \autoref{fig.spectral_perp_long} (left: \(J^\perp(\omega)\), right: \(J^\parallel(\omega)\)) as a function of frequency and \(R_\mathrm{s}\).
We observe that \(J^\parallel(\omega)\) (describing charge-charge interactions) carries almost all the weight, and is approximately given by a Lorentzian line shape whose resonant frequency and width vary with \(R_\mathrm{s}\) due to the sphere-photon coupling.
For \(R_\mathrm{s}\simeq2.5\)~nm, \autoref{fig.spectral_comparison} highlights the large difference in scale, as the peak of \(J^\parallel(\omega)\) is about 5 orders of magnitude larger than \(J^\perp(\omega)\), in agreement with the discussion above.
Additionally, \(J^\perp(\omega)\simeq J_{0}^{\perp}(\omega)\) except for a narrow region around the plasmonic resonance.

It is clear from \autoref{fig.spectral_perp_long} and \autoref{fig.spectral_comparison} that USC can only be achieved through Coulomb interactions.
In \autoref{fig.longitudinality}, we show the ratio between the longitudinal and transverse coupling strengths averaged over the frequency range of the resonance as a function of \(R_\mathrm{s}\) and the emitter-sphere separation.
Its large values (\(>10^5\)) confirm that the emitter interaction is prominently longitudinal even out of the USC region.
The contour lines indicate the fraction of the excitation frequency reached by the longitudinal coupling strength.
Clearly, the system only becomes ultrastrongly coupled due to the longitudinal fields and their enhancement for small \(R_\mathrm{s}\) and emitter-sphere separation.

Last, concerning the PSE term, \autoref{fig.pse_potential} shows the full spatial dependence of \(\phi_\mathrm{s}\) through its linear coefficient.
To lowest order, \(\phi_\mathrm{s}\) depends linearly on \(z_\mathrm{s}\), such that \(\phi_\mathrm{s}(\mathbf{r},z_\mathrm{s}) = K^\parallel(\mathbf{r}) z_\mathrm{s} + \mathcal{O}\left(z_\mathrm{s}^2\right)\).
Here, \(K^\parallel(\mathbf{r})\) is the linear coefficient that encodes the spatial dependence of the potential in the small-oscillation limit (black line).
This potential decays as \(1/|\mathbf{r}|^2\) away from the sphere, and is thus bounded.
The long-wavelength approximation corresponds to replacing this potential by an unbounded linear one (orange line) given by the first-order term in the Taylor series around the emitter position.
Within a large enough spatial box, such a linear potential eventually wins over the molecular binding forces and disintegrates the emitter, as noted in~\cite{rokajNoGroundState2018,schäferAsquared2020}, but as discussed above, that is an artifact introduced by the long-wavelength approximation, and adding a PSE term is the wrong remedy.

In conclusion, we have derived a bound on the maximum possible coupling strength between small emitters and photons.
This bound applies for arbitrary systems and shows that ultrastrong coupling cannot be achieved through the transverse interaction with photons, as it is physically impossible for materials to exist that would achieve the required concentration of transverse fields.
These results have profound implications for theoretical descriptions of light-matter interactions approaching the USC regime, as they imply that most commonly used model Hamiltonians assuming a photon-like cavity mode are not applicable for describing single-emitter USC\@.
In contrast, we find that Coulomb interactions are the dominant mechanism in the USC regime, while transverse photons can be neglected altogether.
The explicit analytically solvable model devised bolsters our claims by explicitly illustrating the argument. Consequently, we hope that this article will inform and clarify the theoretical modeling of ultrastrong light-matter interactions and cavity-modified material properties in future studies.

\bibliography{references}

\section*{Methods}

We provide here the details regarding the analytic example discussed in the main text of an emitter interacting with a spherical plasmonic particle, and both coupled to the photons. We begin with a derivation of the Hamiltonian in \autoref{eq.H_model}. Then, we focus on the sphere-photon subsystem, expanding first their mutual interaction term in a convenient manner. We next show the diagonalization steps, which consist on two successive canonical transformations, followed by a Fano diagonalization. With the diagonalization coefficients, we rewrite the emitter-sphere and emitter-photon interactions in terms of the polaritonic eigenmodes. We finish by providing some technical details regarding \autoref{fig.longitudinality} and \autoref{fig.pse_potential}.

\subsection*{Derivation of the Hamiltonian}

In the model, the emitter is a collection of bound point charges \(q_i\) with small spatial extent and, for concreteness, placed along the \(z\) axis. We represent the plasmonic particle with two overlapping homogeneously charged spheres with radius \(R_\mathrm{s}\) that portray the ionic background (with charge density \(\rho\)) and conduction electrons (\(-\rho\)). The center of the electronic sphere oscillates along the \(z\) axis around the center of the ionic sphere, which remains fixed at the origin. Thus, the dipolar plasmonic resonance is accurately captured in the limit of small oscillation amplitude \(z_\mathrm{s}\). As for the photons, we describe them through \(\mathbf{A}^\perp\), which coincides with the full vector potential in the Coulomb gauge.

We start from the Lagrangian for the full emitter-sphere-photon system:
\begin{align}
    L &= L_\mathrm{e} + L_\mathrm{s} + L_\mathrm{f} + L^\parallel_\mathrm{int, e} + L^\perp_\mathrm{int} \\
    \nonumber
    L_\mathrm{e} &= \sum_{i\in\mathrm{e}} \frac{m_i \dot{\mathbf{r}}_i^2}{2} - \sum_{i > j\in\mathrm{e}} \frac{q_i q_j}{4\pi\varepsilon_0 |\mathbf{r}_i - \mathbf{r}_j|}\\
    \nonumber
    L_\mathrm{s} &= \frac{m_\mathrm{s} \dot{z}_\mathrm{s}^2}{2} + \int\mathrm{d}^3 r\ \rho(\mathbf{r}) \phi_+(\mathbf{r})  - \frac{8\pi\rho^2 R_\mathrm{s}^5}{15\varepsilon_0} \\
    \nonumber
    L_\mathrm{f} &= \frac{\varepsilon_0}{2} \int\mathrm{d}^3 r\ \left[\left(\dot{\mathbf{A}}^\perp(\mathbf{r}) \right)^2 - c^2 \left(\nabla\times\mathbf{A}^\perp(\mathbf{r}) \right)^2\right] \\
    \nonumber
    L^\parallel_\mathrm{int, e} &= - \sum_{i\in\mathrm{e}} q_i \phi_\mathrm{s}(\mathbf{r}_i, z_\mathrm{s}) \\
    \nonumber 
    L^\perp_\mathrm{int} &= \sum_{i\in\mathrm{e}} q_i \dot{\mathbf{r}}_i \cdot \mathbf{A}^\perp(\mathbf{r}_i) + \int \mathrm{d}^3r\ \mathbf{j}_\mathrm{s}(\mathbf{r}) \cdot \mathbf{A}^\perp(\mathbf{r}),
\end{align}
where \(i\) and \(j\) label the emitter's charges, and the divergent Coulomb self-energy of its point charges has been excluded (\(i>j\)). The electronic sphere, with mass \(m_\mathrm{s}\), interacts with the ionic one through the second therm in \(L_\mathrm{s}\), where the charge density is a step function \(\rho(\mathbf{r}) = \rho \theta(R_\mathrm{s} - |\mathbf{r}-\mathbf{z}_\mathrm{s}|)\) and
\begin{equation}\label{eq.phi_plus}
    \phi_+(\mathbf{r}) = \frac{\rho}{3\varepsilon_0}
    \begin{cases}
    \left(3R_\mathrm{s}^2 - |\mathbf{r}|^2\right) / 2 & \mathrm{if}~ |\mathbf{r}| \leq R_\mathrm{s}, \\
    R_\mathrm{s}^3 / |\mathbf{r}| & \mathrm{if}~ |\mathbf{r}| > R_\mathrm{s}.
    \end{cases}
\end{equation}
is the electrostatic potential due to the ionic sphere. Additionally, we include for convenience the finite and constant Coulomb self-energy of the spheres. We write the emitter-sphere interaction in terms of the sphere's Coulomb potential \(\phi_\mathrm{s}\), while the charges couple to the photons through their currents. In particular, \(\mathbf{j}_\mathrm{s}(\mathbf{r}) = -\rho \dot{\mathbf{z}}_\mathrm{s} \theta(R_\mathrm{s} - |\mathbf{r}-\mathbf{z}_\mathrm{s}|)\).

In the limit of small oscillations, it can be shown (see supplementary information for the integration) that
\begin{equation*}
    \int\mathrm{d}^3r\ \rho(\mathbf{r})\phi_+(\mathbf{r}) = \frac{8\pi\rho^2 R_\mathrm{s}^5}{15\varepsilon_0} - \frac{2\pi \rho^2 R_\mathrm{s}^3}{9\varepsilon_0} z_\mathrm{s}^2 + \mathcal{O}\left(|z_\mathrm{s}|^3\right).
\end{equation*}
The first constant cancels the Coulomb self-energy of both spheres, while the quadratic term can be written as
\begin{equation*}
    - \frac{2\pi \rho^2 R_\mathrm{s}^3}{9\varepsilon_0} z_\mathrm{s}^2 = -\frac{m_\mathrm{s}}{2}\frac{\rho e}{3 \varepsilon_0 m_e} z_\mathrm{s}^2 =  -\frac{m_\mathrm{s}}{2}\frac{\Omega_\mathrm{p}^2}{3} z_\mathrm{s}^2,
\end{equation*}
where \(\Omega_\mathrm{p}\) is the plasma frequency of the metal. Note that \(\Omega_\mathrm{p}/\sqrt{3}\) is the dipolar resonance of a plasmonic sphere with a lossless Drude permittivity embedded in free space. The canonical momenta are
\begin{subequations}
    \begin{align}
        \mathbf{p}_i &= m_i \dot{\mathbf{r}}_i + q_i \mathbf{A}^\perp(\mathbf{r}_i) \\
        p_\mathrm{s} &= m_s \dot{z}_\mathrm{s} - \int\mathrm{d}^3 r\ \rho(\mathbf{r})\hat{\mathbf{z}}\cdot \mathbf{A}^\perp(\mathbf{r}) \\
        \boldsymbol{\Pi}(\mathbf{r}) &= \varepsilon_0 \dot{\mathbf{A}}^\perp(\mathbf{r}) = - \varepsilon_0 \mathbf{E}^\perp(\mathbf{r})
    \end{align}
\end{subequations}
and 
\begin{equation}
    H = \sum_i \mathbf{p}_i\cdot \dot{\mathbf{r}}_i + p_\mathrm{s} \dot{z}_\mathrm{s} + \int\mathrm{d}^3r\ \boldsymbol{\Pi}(\mathbf{r})\cdot \dot{\mathbf{A}}^\perp(\mathbf{r}) - L
\end{equation}
yields the Hamiltonian in \autoref{eq.H_model}.

\subsection*{Sphere-photon interactions}

Given the spherical symmetry of the plasmonic particle, it is convenient to expand the fields in the real spherical-wave basis, 
\begin{equation} \label{eq.spherical_expansion}
    \mathbf{A}^\perp(\mathbf{r}) = \sum_{\lambda=\mathrm{te, tm}}\sum_{lm}\int\mathrm{d}\omega\ A_{lm}^{(\lambda)}(\omega) \mathbf{f}_{lm}^{(\lambda)}(\omega, \mathbf{r})
\end{equation}
defined by
\begin{subequations}
    \begin{align}\nonumber
        \mathbf{f}_{lm}^\mathrm{(te)}(\omega, \mathbf{r}) &= \frac{\omega}{c}\sqrt{\frac{2}{l(l+1)\pi c}} j_l\left(\frac{\omega |\mathbf{r}|}{c}\right) \\ 
        &\phantom{=\ space} \left[\hat{\boldsymbol{\vartheta}}\frac{\partial_\varphi}{\sin\vartheta}-\hat{\boldsymbol{\varphi}}\partial_\vartheta\right] Y_{lm}(\vartheta, \varphi)\\
        \mathbf{f}_{lm}^\mathrm{(tm)}(\omega, \mathbf{r}) &= \frac{c}{\omega} \nabla\times \mathbf{f}_{lm}^\mathrm{(te)}(\omega, \mathbf{r}).
    \end{align}
\end{subequations}
Here, \(|\mathbf{r}|\), \(\vartheta\) and \(\varphi\) are spherical coordinates, \(l\geq 1\), \(m=-l, -l+1, \dots, l\), \(j_l\) are spherical Bessel functions, \(Y_{lm}\) are the real spherical harmonics, and \(\omega\in[0,\infty)\). These functions are orthogonal and normalized such that
\begin{align*}
    \int\mathrm{d}^3r\ \mathbf{f}^{(\lambda)}_{lm}(\omega, \mathbf{r}) \cdot \mathbf{f}^{(\lambda')}_{l'm'}(\omega', \mathbf{r}) = \delta_{\lambda\lambda'}\delta_{ll'}\delta_{mm'}\delta(\omega-\omega').
\end{align*}
Then, the integral defining the sphere-photon interaction in \autoref{eq.H_model} is
\begin{equation}
    \int\mathrm{d}^3r\ \rho(\mathbf{r})\hat{\mathbf{z}}\cdot \mathbf{A}^\perp(\mathbf{r}) = \sum_{\lambda lm}\int\mathrm{d}\omega\ A_{lm}^{(\lambda)}(\omega) I_{lm}^{(\lambda)}(\omega),
\end{equation}
with 
\begin{align*}
    I_{lm}^{(\lambda)}(\omega) &= \int\mathrm{d}^3r\ \rho(\mathbf{r})\hat{\mathbf{z}}\cdot\mathbf{f}^{(\lambda)}_{lm}(\omega, \mathbf{r}) \\
    &= \frac{4\rho R_\mathrm{s}^2}{\sqrt{3c}}j_l\left(\frac{\omega R_\mathrm{s}}{c}\right)\delta_{\lambda, \mathrm{tm}}\delta_{l1} \delta_{m0} 
\end{align*}
Reaching the second line requires some lengthy manipulations, explicitly done in the supplementary information. The Kronecker deltas are a consequence of the system's symmetry, which greatly simplifies the sphere-photon coupling, as only the tm, \(l=1\) and \(m=0\) modes are relevant.

\subsection*{Diagonalization steps}

We now focus on the sphere-photon subsystem. In the next steps, we perform two consecutive canonical transformations to prepare the Hamiltonian. Then, the Hamiltonian becomes amenable to the Fano diagonalization procedure, which yields analytic expressions for the diagonalization coefficients. Last, we rewrite the original, uncoupled operators in the eigenmode basis.

As shown above, we only need to include the tm, \(l=1\) and \(m=0\) photon modes, while all the others remain unaffected by the sphere and provide part of the free-space background spectral density for the emitter. Accordingly, we alleviate the notation by dropping the tm, \(l=1\), \(m=0\) indices from the expansion coefficients. Using the basis vector functions' orthonormality, we have
\begin{align} \label{eq.H_sphere-photons}
    \nonumber
    H_\mathrm{s-f} &= \frac{\left[p_\mathrm{s} + \frac{4\rho R_\mathrm{s}^2}{\sqrt{3c}}{\displaystyle\int}\mathrm{d}\omega\ j_1\left(\frac{\omega R_\mathrm{s}}{c}\right)A(\omega)\right]^2}{2m_\mathrm{s}} + \frac{m_\mathrm{s}}{2}\frac{\Omega_\mathrm{p}^2}{3} z_\mathrm{s}^2  \\ 
    &\phantom{=\ } + \int\mathrm{d}\omega\ \left(\frac{\Pi^2(\omega)}{2\varepsilon_0} + \frac{\varepsilon_0}{2}\omega^2 A^2(\omega)\right),
\end{align}
where \(\Pi(\omega)\) is the expansion coefficient of \(\boldsymbol{\Pi}\) analogous to \(A(\omega)\) in \autoref{eq.spherical_expansion}. Fundamentally, \autoref{eq.H_sphere-photons} describes a set of interacting harmonic oscillators, coupled through \(p_\mathrm{s}A(\omega)\) and \(A(\omega)A(\omega')\) terms. We may remove the cross-coupling between different \(A(\omega)\) by switching to the multipolar coupling picture with a canonical transformation that replaces the current canonical momenta with
\begin{subequations}
    \begin{align}
        p_\mathrm{s} &\rightarrow p'_\mathrm{s} = p_\mathrm{s} + \frac{4\rho R_\mathrm{s}^2}{\sqrt{3c}}\int\mathrm{d}\omega\ j_1\left(\frac{\omega R_\mathrm{s}}{c}\right)A(\omega) \\
        \Pi(\omega) &\rightarrow \Pi'(\omega) = \Pi(\omega) + \frac{4\rho R_\mathrm{s}^2}{\sqrt{3c}} j_1\left(\frac{\omega R_\mathrm{s}}{c}\right) z_\mathrm{s}.
    \end{align}
\end{subequations}
Note that this is equivalent to adding to the Lagrangian the following total temporal derivative
\begin{equation}
    \frac{\mathrm{d}}{\mathrm{d}t}\left[\frac{4\rho R_\mathrm{s}^2}{\sqrt{3c}}z_\mathrm{s}\int\mathrm{d}\omega\ j_1\left(\frac{\omega R_\mathrm{s}}{c}\right)A(\omega) \right],
\end{equation}
which of course leaves Hamilton's equations invariant. Finally, the sphere-photon Hamiltonian is
\begin{align} \label{eq.H_sphere-photons_multipolar}
    \nonumber
    H_\mathrm{s-f} &= \frac{\left(p'_\mathrm{s}\right)^2}{2m_\mathrm{s}} +\frac{m_\mathrm{s}}{2} \left[\frac{\Omega_\mathrm{p}^2}{3} + \frac{16\rho^2R_\mathrm{s}^4}{3\varepsilon_0 m_\mathrm{s}c} \int\mathrm{d}\omega\ j_1^2\left(\frac{\omega R_\mathrm{s}}{c}\right) \right] z_\mathrm{s}^2\\ 
    \nonumber
    &\phantom{=\ } + \int\mathrm{d}\omega\ \left(\frac{\left(\Pi'(\omega)\right)^2}{2\varepsilon_0} + \frac{\varepsilon_0}{2}\omega^2 A^2(\omega)\right) \\
    &\phantom{=\ } - z_\mathrm{s}\int\mathrm{d}\omega\ \frac{4\rho R_\mathrm{s}^2}{\varepsilon_0\sqrt{3c}} j_1\left(\frac{\omega R_\mathrm{s}}{c}\right)\Pi'(\omega).
\end{align}
In the first line, a finite PSE contribution, equal to \(2\Omega_\mathrm{p}^2/3\) (shown in the supplementary information), is added to the sphere's oscillation frequency. Nevertheless, the actual oscillation frequency is still almost \(\Omega_\mathrm{p}/\sqrt{3}\), as seen in \autoref{fig.spectral_perp_long}. In essence, the multipolar coupling to the photons effectively induces a larger frequency shift on the sphere that compensates the PSE renormalization from the first line above. Although this Hamiltonian has no photonic mode cross-couplings, it is still not in a form that can be easily diagonalized because the interaction involves a position and a momentum.
\begin{subequations}
The problem can be remedied with another canonical transformation, namely
    \begin{align}
        p'_\mathrm{s} &\rightarrow p = \frac{1}{\sqrt{m_\mathrm{s}}}p'_\mathrm{s} && (p\ \text{to}\ p) \\
        z_\mathrm{s} &\rightarrow z = \sqrt{m_\mathrm{s}}z_\mathrm{s}  && (x\ \text{to}\ x) \\
        \Pi'(\omega) &\rightarrow X(\omega) = - \frac{1}{\sqrt{\varepsilon_0}\omega} \Pi'(\omega) && (p\ \text{to}\ -x)\\
        A(\omega) &\rightarrow P(\omega) = \sqrt{\varepsilon_0}\omega A(\omega) && (x\ \text{to}\ p), 
    \end{align}
\end{subequations}
through which the roles of field positions and momenta are swapped, and the coordinates are all scaled for convenience. Now, the sphere-photon Hamiltonian is
\begin{align}\label{eq.H_sphere-photons_preFano}
    \nonumber
    H_\mathrm{s-f} &= \frac{1}{2}\left[p^2 + \Omega_\mathrm{p}^2 z^2 + \int\mathrm{d}\omega \left(P^2(\omega) + \omega^2 X^2(\omega)\right)\right] \\
    &\phantom{=\ } + z \int\mathrm{d}\omega\ \gamma(\omega) X(\omega),
\end{align}
where the coupling strength is
\begin{equation}
    \gamma(\omega) = 2\Omega_\mathrm{p}\sqrt{\frac{R_\mathrm{s}}{\pi c}} \omega j_1\left(\frac{\omega R_\mathrm{s}}{c}\right),
\end{equation}
and the interaction only involves positions (\(z\) and \(X(\omega)\)). Here, we finally apply the Fano diagonalization procedure, briefly outlined in the following. 
\begin{subequations}
    We define new eigenmode coordinates as
    \begin{align}
        \beta(\Omega) &= c_1(\Omega)z + \int\mathrm{d}\omega\ c_2(\Omega, \omega) X(\omega) && (x)\\
        \xi(\Omega) &= c_1(\Omega)p + \int\mathrm{d}\omega\ c_2(\Omega, \omega) P(\omega) && (p).
    \end{align}
\end{subequations}
Then, the goal is to find \(c_1(\omega)\) and \(c_2(\Omega, \omega)\), which can be understood as the matrix coefficients of the orthogonal transformation that diagonalizes the quadratic form of the position sector in the Hamiltonian. From \([H_\mathrm{s-f}, \xi(\Omega)]=i\hbar\Omega^2\beta(\Omega)\), the coefficients fulfill
\begin{subequations}
    \begin{align}
        c_1(\Omega)\left(\Omega^2-\Omega_\mathrm{p}^2\right) &= \int\mathrm{d}\omega\ c_2(\Omega, \omega)\gamma(\omega)\\
        c_2(\Omega, \omega)\left(\Omega^2-\omega^2\right) &= \gamma(\omega)c_1(\Omega).
    \end{align}
\end{subequations}
These can be formally solved by
\begin{subequations}
    \begin{align}
        \nonumber
        c_2(\Omega&,\omega) = \bigg[\mathrm{PV}\frac{1}{\Omega^2-\omega^2} \\
        &+ \frac{\Omega^2-\Omega_\mathrm{p}^2-F(\Omega)}{\gamma^2(\Omega)}\delta(\Omega-\omega)\bigg]\gamma(\omega) c_1(\Omega),
    \end{align}
    through which \(c_2\) depends on \(c_1\). The PV denotes the principal value integral, and
    \begin{align}
        \nonumber
        F(\Omega) &= \mathrm{PV}\int\mathrm{d}\omega\ \frac{\gamma^2(\omega)}{\Omega^2-\omega^2} \\
        &= 2\Omega_\mathrm{p}^2 \frac{\Omega R_\mathrm{s}}{c} j_1\left(\frac{\Omega R_\mathrm{s}}{c}\right) y_1\left(\frac{\Omega R_\mathrm{s}}{c}\right)
    \end{align}
    is an energy shift that compensates the PSE-induced frequency renormalization. The functions \(j_1\) and \(y_1\) are the first spherical Bessel functions of the first and second kind, respectively.
\end{subequations}
Finally, imposing the normalization from \([\beta(\Omega), \xi(\Omega')]=i\hbar\delta\left(\Omega-\Omega'\right)\), we arrive at
\begin{equation}
    c_1(\Omega) = \frac{\gamma(\Omega)}{\sqrt{(\Omega^2-\Omega_\mathrm{p}^2-F(\Omega))^2 + \left(\frac{\pi}{2\Omega}\right)^2 \gamma^4(\Omega)}}
\end{equation} 
after some manipulations. For the remainder of the methods, we require the inverse transformation to express the interactions in terms of the polaritonic eigenmodes. Because the transformation is orthogonal, its inverse is given by the same coefficients as follows:
\begin{subequations}
    \begin{align}
        z_\mathrm{s} &= \frac{1}{\sqrt{m}_\mathrm{s}}\int\mathrm{d}\Omega\ c_1(\Omega) \beta(\Omega)\\
        p'_\mathrm{s} &= \sqrt{m}_\mathrm{s}\int\mathrm{d}\Omega\ c_1(\Omega) \xi(\Omega)\\
        A(\omega) &= \frac{1}{\sqrt{\varepsilon_0}\omega}\int\mathrm{d}\Omega\ c_2(\Omega, \omega) \xi(\Omega)\\
        \Pi'(\omega) &= -\sqrt{\varepsilon_0}\omega\int\mathrm{d}\Omega\ c_2(\Omega, \omega) \beta(\Omega).
    \end{align}
\end{subequations}

\subsection*{Rewriting the interactions}

It is now a straightforward task to rewrite the interaction of the emitter with the sphere and photons in terms of the new eigenmodes. First, the emitter-sphere Coulomb coupling for an emitter placed along the \(z\) axis is
\begin{align}
    \nonumber
    H^\parallel_\mathrm{int,e} &= \sum_{i\in\mathrm{e}} q_i \phi_\mathrm{s}(\mathbf{r}_i, z_\mathrm{s}) \simeq -z_\mathrm{s}\sum_{i\in\mathrm{e}} \frac{q_i\rho  R_\mathrm{s}^3}{3\varepsilon_0|\mathbf{r}_i|^3}\hat{\mathbf{z}}\cdot \mathbf{r}_i \\
    \nonumber
    &= \sum_{i\in\mathrm{e}} \frac{q_i\rho  R_\mathrm{s}^3}{3\varepsilon_0\sqrt{m_\mathrm{s}}|\mathbf{r}_i|^3}\hat{\mathbf{z}}\cdot \mathbf{r}_i \int\mathrm{d}\Omega\ c_1(\Omega) \beta(\Omega) \\
    \nonumber
    &\simeq \frac{2\rho R_\mathrm{s}^3 \mathbf{d}\cdot\hat{\mathbf{z}}}{3\varepsilon_0\sqrt{m_\mathrm{s}}|\mathbf{r}_\mathrm{e}|^3}\int\mathrm{d}\Omega\ c_1(\Omega) \sqrt{\frac{\hbar}{2\Omega}}\left(b^\dagger(\Omega) + b(\Omega)\right) \\
    &= \hbar\sum_t \left(\sigma_t^\dagger + \sigma_t\right)\int\mathrm{d}\Omega\ g_t^\parallel(\Omega) \left(b^\dagger(\Omega) + b(\Omega)\right)
\end{align}
In the first line, we have used the small-oscillation approximation to write a compact expression for \(\phi_\mathrm{s}(\mathbf{r}_i, z_\mathrm{s})\). Additionally, we have done the long-wavelength approximation in the second-to-last equation, with the emitter's dipole moment \(\mathbf{d}\) and position \(\mathbf{r}_\mathrm{e}\). The last line splits the dipole moment in transitions \(t\) as in the main text, and collects most factors into \(g_t^\parallel\). From it, the longitudinal spectral density \(J_t^\parallel\) represented in the right panel of \autoref{fig.spectral_perp_long} is defined as
\begin{equation}
    J_t^\parallel(\Omega) = \left(g_t^\parallel(\Omega)\right)^2 = \frac{ |\mathbf{d}_t|^2\Omega_\mathrm{p}^2 R_\mathrm{s}^3}{36\pi \hbar\Omega \varepsilon_0 |\mathbf{r}_\mathrm{e}|^6}c_1^2(\Omega).
\end{equation}

Next, we consider the emitter-photon interaction, which requires further manipulations. Similar to \autoref{eq.H_int_transverse_photon} in the main text, we have
\begin{align}
    \nonumber
    H_\mathrm{int,e}^{\perp} &= - \sum_{i\in\mathrm{e}} \frac{q_i}{m_i}\mathbf{p}_i\cdot \sum_\lambda\sum_{lm}\int\mathrm{d}\omega\ A_{lm}^{(\lambda)}(\omega) \mathbf{f}_{lm}^{(\lambda)}(\omega, \mathbf{r}_i).
\end{align}
As mentioned before, only the tm, \(l=1\), \(m=0\) modes interact with the sphere and give rise to polaritonic eigenmodes. For that reason, let us consider here only the contribution due to these modes and leave the rest for a short comment below. Then, when \(\mathbf{d}_t \parallel \hat{\mathbf{z}}\),
\begin{align}
    \nonumber
    H_\mathrm{int,e}^{\perp} &= - \sum_{i\in\mathrm{e}} \frac{q_i}{m_i}\mathbf{p}_i\cdot \int\mathrm{d}\omega\ A(\omega) \mathbf{f}_{10}^\mathrm{(tm)}(\omega, \mathbf{r}_i) \\
    \simeq -\hbar&\sum_t \left(\sigma_t-\sigma_t^\dagger\right) \int\mathrm{d}\Omega\ \frac{\omega_t}{\Omega} g_t^\perp(\Omega) \left(b^\dagger(\Omega) - b(\Omega)\right),
\end{align}
where 
\begin{align}
    g_t^\perp(\Omega) &= |\mathbf{d}_t|\sqrt{ \frac{\Omega^3}{2\hbar\varepsilon_0}} \int\mathrm{d}\omega\ \frac{c_2(\Omega, \omega) \hat{\mathbf{z}}\cdot\mathbf{f}_{10}^\mathrm{(tm)}(\omega, \mathbf{r}_\mathrm{e})}{\omega}.
\end{align}
Fortunately, the integral above can be solved analytically:
\begin{widetext}
    \begin{equation}
        \label{eq.meth_g_perp} 
        g_t^\perp(\Omega) = \frac{|\mathbf{d}_t|}{|\mathbf{r}_\mathrm{e}|}\sqrt{\frac{3\Omega^3}{2\pi^2\hbar\varepsilon_0 c}} c_1(\Omega)\left\{\frac{\big(\Omega^2-\Omega_\mathrm{p}^2-F(\Omega)\big)j_1\Big(\frac{\Omega R_\mathrm{s}}{c}\Big)}{\Omega \gamma(\Omega)} + \Omega_\mathrm{p}\frac{\sqrt{\pi c R_\mathrm{s}^3}}{|\mathbf{r}_\mathrm{e}|^2 \Omega^2} \left[\frac{1}{3} + \frac{\Omega|\mathbf{r}_\mathrm{e}|^2 j_1\Big(\frac{\Omega R_\mathrm{s}}{c}\Big) y_1 \Big(\frac{\Omega|\mathbf{r}_\mathrm{e}|}{c}\Big) }{c R_\mathrm{s}}\right]\right\},
    \end{equation}
\end{widetext}
from which the transverse spectral density \(J_t^\perp\) is
\begin{equation}
    J_t^\perp(\Omega) = \left(g_t^\perp(\Omega)\right)^2.
\end{equation}
We show the transverse spectral density on the left panel in \autoref{fig.spectral_perp_long}, which also accounts for the rest of the free-space modes discussed next.

\subsubsection*{Free-space modes}

The photonic modes that do not interact with the sphere remain as a slightly modified free-space contribution to the emitter. Accordingly, these modes induce a Lamb shift and spontaneous emission rate, much smaller than the effects due to the sphere-photon eigenmodes. After using the form of the real spherical-wave functions and manipulations similar to the ones above, we obtain for these modes
\begin{widetext}
    \begin{align}
        \nonumber
        H_\mathrm{int,e}^{\perp,0} &= - \sum_{i\in\mathrm{e}} \frac{q_i}{m_i}\mathbf{p}_i\cdot \sum_{l>1}\int\mathrm{d}\omega\ A_{l0}^\mathrm{(tm)}(\omega) \mathbf{f}_{l0}^\mathrm{(tm)}(\mathbf{r}_i) \\
        \nonumber
        &\simeq i\hbar\sum_t \left(\sigma_t-\sigma_t^\dagger\right) \int\mathrm{d}\omega\ \frac{\omega_t}{\omega |\mathbf{r}_\mathrm{e}|} \sqrt{\frac{|\mathbf{d}_t|^2\omega}{4\pi^2\hbar\varepsilon_0c}}\sum_{l>1} \sqrt{l(l+1)(l+2)} j_l\left(\frac{\omega|\mathbf{r}_\mathrm{e}|}{c}\right)
        \left(a_{l0}^{\mathrm{(tm)}}(\omega) + \left(a_{l0}^{\mathrm{(tm)}}(\omega)\right)^\dagger\right)\\
        \nonumber
        &= i\hbar\sum_t \left(\sigma_t-\sigma_t^\dagger\right)  \int\mathrm{d}\omega\ \frac{\omega_t}{\omega |\mathbf{r}_\mathrm{e}|} \sqrt{\frac{|\mathbf{d}_t|^2\omega}{4\pi^2\hbar\varepsilon_0c}}\sqrt{\sum_{l>1} l(l+1)(l+2) j_l^2\left(\frac{\omega|\mathbf{r}_\mathrm{e}|}{c}\right)}
        \left(c(\Omega) + c^\dagger(\Omega)\right)\\
        &= i\hbar\sum_t \left(\sigma_t-\sigma_t^\dagger\right) \int\mathrm{d}\omega\ \frac{\omega_t}{\omega} \sqrt{J_{t,0}^{\perp}(\omega)\left[1 - \left(\frac{3j_1\left(\frac{\omega |\mathbf{r}_\mathrm{e}|}{c}\right)}{\frac{\omega |\mathbf{r}_\mathrm{e}|}{c}}\right)^2\right]}
        \left(c(\Omega) + c^\dagger(\Omega)\right).
    \end{align}
\end{widetext}
Here, we first expand the relevant part of the mode functions. Then, we unitarily combine all the modes with the same frequency into 
\begin{equation}
    c(\Omega) = \frac{\sum_{l>1} \sqrt{l(l+1)(l+2)} j_1\left(\frac{\omega|\mathbf{r}_\mathrm{e}|}{c}\right) a_{l0}^\mathrm{(tm)}(\omega)}{\sqrt{\sum_{l>1} l(l+1)(l+2) j_1^2\left(\frac{\omega|\mathbf{r}_\mathrm{e}|}{c}\right)}}.
\end{equation}
Now, we use that
\begin{equation}
    3\sum_{l} l(l+1)(l+2)j_l^2(x) = 2x^2
\end{equation}
to recover the free-space spectral density
\begin{equation}
    J_{t,0}^{\perp}(\omega) = \frac{|\mathbf{d}_t|^2\omega^3}{6\pi^2\hbar\varepsilon_0c^3},
\end{equation}
modified by the factor in the square brackets that suppresses it for \(\omega \ll c/|\mathbf{r}_\mathrm{e}|\).

Last, we remark that \(c_2(\Omega, \omega) \rightarrow -\delta(\Omega-\omega)\)
far from the resonance. This limit implies that the high-frequency polaritonic eigenmodes are essentially photon modes, as the sphere is unable to respond at such high frequencies. From \autoref{fig.spectral_perp_long} and \autoref{fig.spectral_comparison}, it is clear that \(\Omega_\mathrm{p}\) is already high enough and, accordingly, even the tm, \(l=1\), \(m=0\) photons behave as in free-space for \(\omega>\Omega_\mathrm{p}\).

\subsection*{Polaritonic PSE}

We calculate here the relevant part of the PSE, and show that it is negligible anyway. The full PSE involves a sum over all the photonic modes that is independent of the nanostructure. This quantity is formally divergent, and perturbative treatments show that it contributes in the renormalization of the emitter's free-space Lamb shift. However, we are treating here a subset of the photonic modes explicitly, namely, those that interact with the sphere (tm, \(l=1\), \(m=0\)). From these modes, we have additionally seen above that only those with \(\omega<\Omega_\mathrm{p}\) can really interact with the sphere. Consequently, we only need to include explicitly the tm, \(l=1\), \(m=0\) modes with \(\omega<\Omega_\mathrm{p}\), while the rest can be accounted for perturbatively by including a very small Lamb shift and spontaneous emission rate. Then, the part of the PSE associated to the tm, \(l=1\), \(m=0\) photons with \(\omega<\Omega_\mathrm{p}\) should be present in the multipolar coupling Hamiltonian:
\begin{align}
    \nonumber
    H_\mathrm{PSE}^\mathrm{expl.} &\simeq \hbar|\hat{\mathbf{z}}\cdot\mathbf{d}|^2 \int_0^{\Omega_\mathrm{p}}\mathrm{d}\omega\ \frac{3j_1^2\left(\frac{\omega|\mathbf{r}_\mathrm{e}|}{c}\right)}{2\hbar\varepsilon_0\pi^2c|\mathbf{r}_\mathrm{e}|^2} \simeq \frac{|\hat{\mathbf{z}}\cdot\mathbf{d}|^2\Omega_\mathrm{p}^3}{18\varepsilon_0\pi^2c^3}\\
    &\simeq 10^{-5}~ \mathrm{eV}/(e\cdot\mathrm{nm})^2 \times |\hat{\mathbf{z}}\cdot\mathbf{d}|^2 ,
\end{align}
where we have expanded the transverse part of the emitter's polarization,
\begin{equation}
    \mathbf{P}_\mathrm{e}(\mathbf{r}) = \sum_{i\in\mathrm{e}} q_i \mathbf{r}_i \int_0^1\mathrm{d}\sigma\ \delta(\mathbf{r}-\sigma\mathbf{r}_i)
\end{equation}
in the spherical-wave basis and retained only the tm, \(l=1\), \(m=0\) coefficients up to \(\omega = \Omega_\mathrm{p}\). The first approximate equality comes from the long-wavelength approximation, while the second relies on \(|\mathbf{r}_\mathrm{e}|\ll c / \Omega_\mathrm{p}\simeq 20\)~nm (for \(\rho=60~e\cdot\mathrm{nm}^{-3}\)). Numerically evaluating the result leads to the conclusion that the explicit part of the PSE, although formally present in the Hamiltonian, is completely negligible due to the small prefactor. Before moving on to the next subsection, we note that the integral above is analytical, and given in the supplementary information.

\subsection*{Details for \autoref{fig.longitudinality}}

We have defined the longitudinality \(\ell\) as 
\begin{equation}
    \ell = \frac{\left\langle J^\parallel \right\rangle _\mathrm{res.}}{\left\langle J^\perp \right\rangle _\mathrm{res.}},
\end{equation}
where \(\langle\cdots\rangle_\mathrm{res.}\) denotes that the average is taken over a narrow frequency interval covering the resonance, \(\left(0.57\Omega_\mathrm{p}, 0.58\Omega_\mathrm{p}\right)\). Accordingly, large values correspond to a dominant longitudinal interaction. The contour lines in the figure quantify the USC parameter. To calculate it, we fit \(J^\parallel\) to a Lorentzian function
\begin{equation}
    L(\omega) = \frac{\left(G_\mathrm{res.}^\parallel\right)^2}{\pi} \frac{\kappa_\mathrm{res.}/2}{(\omega-\omega_\mathrm{res.})^2 + (\kappa_\mathrm{res.}/2)^2},
\end{equation}
and use that a Lorentzian spectral density is equivalent to a single lossy mode, with coupling strength \(G_\mathrm{res.}^\parallel\). Then, the contours show the value of \(\frac{d G^\parallel_\mathrm{res.}}{\omega_\mathrm{res.}}\), where \(d=10\)~Debye is the transition dipole moment.

\subsection*{Details for \autoref{fig.pse_potential}}

The potential due to the sphere can be extracted from the potential of the individual spheres. Recalling \autoref{eq.phi_plus}, the total potential is simply
\begin{align}
    \nonumber
    \phi_\mathrm{s}(\mathbf{r}, z_\mathrm{s}) &= \phi_+(\mathbf{r}) + \phi_-(\mathbf{r}) = \phi_+(\mathbf{r}) - \phi_+(\mathbf{r}-z_\mathrm{s}\hat{\mathbf{z}})\\ 
    \nonumber
    &= z_\mathrm{s}\hat{\mathbf{z}}\cdot \nabla \phi_+(\mathbf{\mathbf{r}}) + \mathcal{O}(z_\mathrm{s}^2) \\
    &\simeq z_\mathrm{s}\frac{\rho \hat{\mathbf{z}}\cdot\mathbf{r}}{3\varepsilon_0} \begin{cases}
        -1 & \mathrm{if}~ |\mathbf{r}| \leq R_\mathrm{s} \\
        -R_\mathrm{s}^3 / |\mathbf{r}|^3 & \mathrm{if}~ |\mathbf{r}| > R_\mathrm{s}
    \end{cases} = K^\parallel z_\mathrm{s}.
\end{align}
Dropping the higher order terms, we see that \(\phi_\mathrm{s}\) is linear in \(\mathbf{r}\) when \(|\mathbf{r}|\leq R_\mathrm{s}\) and essentially recovers the potential due to a dipole outside the sphere.

\end{document}


\title{Supplementary Information for: There is no ultrastrong coupling with photons}

\author{Diego Fernández de la Pradilla}
    \email{diego.fernandez@uam.es}
    \affiliation{Departamento de Física Teórica de la Materia Condensada and Condensed Matter Physics Center (IFIMAC), Universidad Autónoma de Madrid, E-28049 Madrid, Spain}
\author{Esteban Moreno}%
    \email{esteban.moreno@uam.es}
    \affiliation{Departamento de Física Teórica de la Materia Condensada and Condensed Matter Physics Center (IFIMAC), Universidad Autónoma de Madrid, E-28049 Madrid, Spain}
\author{Johannes Feist}%
    \email{johannes.feist@uam.es}
    \affiliation{Departamento de Física Teórica de la Materia Condensada and Condensed Matter Physics Center (IFIMAC), Universidad Autónoma de Madrid, E-28049 Madrid, Spain}
\date{\today}

\maketitle

\allowdisplaybreaks

This supplementary information contains a proof of Eqs.~(9) and (14) of the main text. The first one is the sum rule used to derive the bound on the transverse coupling strength, and the second one relates the PSE to the emitter-polariton coupling strengths. Next, we provide practical information concerning the analytical evaluation of integrals and particularly involved calculation steps of the explicit analytical model. In particular, we calculate here (i) the Coulomb interaction between the spheres, (ii) the integral in the sphere-photon interaction term, (iii) the sphere's frequency shift in the multipolar coupling scheme, (iv) the function \(F(\Omega)\), (v) the coefficient \(c_1(\Omega)\), (vi) the emitter-photon interaction in the polaritonic eigenbasis, and (vii) emitter's self-energy due to the ``true'' polaritonic modes, i.e., those that behave differently from the free-space case. Last, we calculate the total spectral density including both longitudinal and transverse fields by moving into a complete multipolar picture, which allows us to compare with the spectral density as obtained in the framework of macroscopic QED.

\section*{Proof of the electromagnetic sum rule}

The sum rule in Eq.~(9) of the main text relates the transverse spectral density \(J_t^\perp\) and the free-space spectral density \(J_{t,0}^{\perp}\), which determine the transverse interaction strength of the emitter to the polaritonic (cavity-photon) modes and to the photonic modes, respectively. To properly identify each spectral density, let us begin from the minimal coupling, Coulomb gauge Hamiltonian for an emitter, a cavity, and the transverse EM field:
\begin{align}
    \nonumber
    H =&\, 
    \underset{H_\mathrm{e}}{\underbrace{\sum_{i\in\mathrm{e}} \frac{\mathbf{p}_i^2}{2m_i} + \sum_{i>j\in\mathrm{e}} \frac{q_i q_j}{4\pi\varepsilon_0 \left|\mathbf{r}_i - \mathbf{r}_j\right|}}} 
    + \underset{H_\mathrm{c}}{\underbrace{\sum_{\alpha\alpha'} \left[\frac{\delta_{\alpha\alpha'}}{2}P_\alpha P_{\alpha'} + \frac{\omega_\alpha^2 \delta_{\alpha\alpha'} + c_{\alpha\alpha'}}{2} X_\alpha X_{\alpha'}\right]}} 
    + \underset{H_\mathrm{int,e}^\parallel}{\underbrace{\sum_\alpha\int\mathrm{d}^3r\ \mathbf{P}_\mathrm{e}(\mathbf{r}) \cdot \bE^\parallel_\alpha(\mathbf{r}) X_\alpha }}\\
    \nonumber
    &\, 
    + \underset{H_\mathrm{f}}{\underbrace{\sum_{\lambda}\left[\frac{1}{2\varepsilon_0}\Pi_\lambda^2 + \frac{\varepsilon_0 \omega_\lambda^2 }{2} A^2_\lambda\right]}} 
    + \underset{H^{\perp}_\mathrm{int, c}}{\underbrace{\sum_{\alpha\lambda} C_{\alpha\lambda}P_\alpha A_\lambda}} 
    - \underset{-H^{\perp}_\mathrm{int, e}}{\underbrace{\sum_{i\in\mathrm{e}} \frac{q_i}{m_i} \mathbf{p}_i\cdot \sum_\lambda A_\lambda \mathbf{f}_\lambda(\mathbf{r}_i) }}\\
    &\,
    \label{eq.Hamiltonian_minimalcoupling}
    + \underset{H_\mathrm{diam,c}}{\underbrace{\sum_{\lambda\lambda'} \sum_\alpha \frac{C_{\alpha\lambda} C_{\alpha\lambda'}}{2} A_\lambda A_{\lambda'}}} 
    + \underset{H_\mathrm{diam,e}}{\underbrace{\sum_i \frac{q_i^2}{2m_i} \left(\sum_\lambda A_\lambda \mathbf{f}_\lambda(\mathbf{r}_i)\right)^2}}.
\end{align}
In the above Hamiltonian, we assume that the cavity's response is linear, and can thus be modelled as some set of polarization harmonic oscillators. Their coordinates and momenta are denoted by \(P_\alpha\) and \(X_\alpha\), and we allow for the presence of some interaction between them through \(c_{\alpha\alpha'}\). Note that the plasmon model discussed in the main text is a particularly simple instance of this more general scenario, in which there is only one polarization degree of freedom \(\alpha\), associated to the dipolar plasmonic mode (\(p_\mathrm{s}, z_\mathrm{s}\)). The third term corresponds to the longitudinal Coulomb interaction between the emitter, through its polarization density \(\mathbf{P}_\mathrm{e}\), and the field due to each polarization mode in the cavity \(\bE^\parallel_\alpha X_\alpha\). The second line starts with the field Hamiltonian, written as a sum over photonic modes, whose canonical coordinates and momenta are \(A_\lambda\) and \(\Pi_\lambda\), defined through
\begin{equation}
    \mathbf{A}^\perp(\mathbf{r}) = \sum_\lambda A_\lambda \mathbf{f}_\lambda(\mathbf{r}) 
    \hspace{0.3cm}\text{and}\hspace{0.3cm} 
    -\varepsilon_0\mathbf{E}^\perp(\mathbf{r}) = \sum_\lambda \Pi_\lambda \mathbf{f}_\lambda(\mathbf{r}),
\end{equation}
respectively, where \(\mathbf{f}_\lambda\) are the mode functions that span the space of transverse vector fields. Next, we have the transverse interaction of the photonic modes with the cavity and the emitter. In \(H_\mathrm{int, c}^{\perp}\), we have kept the coupling \(C_{\alpha\lambda}\) unspecified because it is not relevant for the argument to be explained below, but it physically must be a spatial average of \(\mathbf{A}^\perp(\mathbf{r})\) weighted by the charge distribution profile of the polarization mode. Again, a simple version can be found in the Hamiltonian for the plasmonic model, Eq.~(17) of the main text. Finally, the last line includes the diamagnetic \(A^2\) terms due to the cavity and the emitter. 

From \(H_\mathrm{int, e}^{\perp}\) in  \autoref{eq.Hamiltonian_minimalcoupling}, we can already write an expression for the free-space spectral density \(J_{t,0}^{\perp}\), as done in the main text [Eq.~(8)]. Indeed, using the relation between momentum and position matrix elements \(\mathrm{Tr}\left\{\sigma_t^\dagger \mathbf{p}_i \right\} = -i \omega_t m_i \mathrm{Tr}\left\{\sigma_t^\dagger \mathbf{r}_i\right\}\) and introducing the photonic ladder operators through \(A_\lambda = \sqrt{\frac{\hbar}{2\varepsilon_0 \omega_\lambda}}\left(a^\dagger_\lambda + a_\lambda\right)\), we can rewrite the emitter-field interaction in the long-wavelength approximation as
\begin{equation}
    H_\mathrm{int, e}^{\perp} \approx i\hbar \sum_{t\lambda} \sigma_t \frac{\omega_t}{\omega_\lambda} \sqrt{\frac{\omega_\lambda}{2\hbar\varepsilon_0}} \mathbf{d}_t \cdot \mathbf{f}_\lambda(\mathbf{r}_\mathrm{e})
    \left(a^\dagger_\lambda + a_\lambda\right) + \mathrm{H.c.} = i\hbar \sum_{t\lambda} \sigma_t \frac{\omega_t}{\omega_\lambda} g_{t\lambda,0}^{\perp}
    \left(a^\dagger_\lambda + a_\lambda\right) + \mathrm{H.c.},
\end{equation}
where
\begin{equation}
    \label{eq.minimal_g0}
    g_{t\lambda,0}^{\perp} = \sqrt{\frac{\omega_\lambda}{2\hbar\varepsilon_0}}\mathbf{d}_t\cdot \mathbf{f}_\lambda(\mathbf{r}_\mathrm{e}).
\end{equation}
Thus, the transverse spectral density that characterizes the emitter-photon coupling strength is
\begin{equation}
    J_{t, 0}^{\perp}(\omega) = \sum_\lambda \left(g_{t\lambda,0}^{\perp}\right)^2 \delta(\omega-\omega_\lambda) = \sum_\lambda \frac{\omega_\lambda}{2\hbar\varepsilon_0}\left[\mathbf{d}_t\cdot \mathbf{f}_\lambda(\mathbf{r}_\mathrm{e})\right]^2 \delta(\omega-\omega_\lambda), 
\end{equation}
that is, the total interaction strength between the emitter transition \(\sigma_t\) and all the modes \(\lambda\) with a given frequency \(\omega\). An elementary evaluation of the sum for purely photonic modes (free-space) yields the well-known result of \( J_{t, 0}^{\perp}(\omega) = \frac{\left|\mathbf{d}_t\right|^2\omega^3}{6\pi^2\hbar\varepsilon_0c^3}\). Once \(J_{t, 0}^{\perp}\) is known, in order to derive the sum rule we must find an expression for \(J_t^\perp\).

Our next step involves diagonalizing the cavity-photon subsystem and expressing \(H_\mathrm{int,\mathrm{e}}^{\perp}\) in terms of the new polaritonic eigenmodes. To that end, we first perform a canonical transformation of the field and cavity momenta:
\begin{align}
    P_\alpha \rightarrow P'_\alpha = P_\alpha + \sum_\lambda C_{\alpha\lambda} A_\lambda 
    \hspace{0.3cm}\text{and}\hspace{0.3cm}
    \Pi_\lambda \rightarrow \Pi'_\lambda = \Pi_\lambda + \sum_\alpha C_{\alpha\lambda} X_\alpha,
\end{align}
which effectively switches to a multipolar coupling picture. In terms of the new momenta, the Hamiltonian in \autoref{eq.Hamiltonian_minimalcoupling} becomes
\begin{align}
    \nonumber
    H =&\, 
    \underset{H_\mathrm{e}}{\underbrace{\sum_{i\in\mathrm{e}} \frac{\mathbf{p}_i^2}{2m_i} + \sum_{i>j\in\mathrm{e}} \frac{q_i q_j}{4\pi\varepsilon_0 \left|\mathbf{r}_i - \mathbf{r}_j\right|}}} 
    + \underset{H_\mathrm{c}}{\underbrace{\sum_{\alpha\alpha'} \left[\frac{\delta_{\alpha\alpha'}}{2}P_\alpha' P_{\alpha'}' + \frac{\omega_\alpha^2 \delta_{\alpha\alpha'} + c_{\alpha\alpha'}}{2} X_\alpha X_{\alpha'}\right]}} 
    + \underset{H_\mathrm{int,e}^\parallel}{\underbrace{\sum_\alpha\int\mathrm{d}^3r\ \mathbf{P}_\mathrm{e}(\mathbf{r}) \cdot \bE^\parallel_\alpha(\mathbf{r}) X_\alpha }}\\
    \nonumber
    &\, 
    + \underset{H_\mathrm{f}}{\underbrace{\sum_{\lambda}\left[\frac{1}{2\varepsilon_0}\left(\Pi'_\lambda\right)^2 + \frac{\varepsilon_0 \omega_\lambda^2 }{2} A^2_\lambda\right]}} 
    - \underset{-H_\mathrm{int, c}^{\perp}}{\underbrace{\sum_{\alpha\lambda} \frac{C_{\alpha\lambda}}{2\varepsilon_0}X_\alpha \Pi'_\lambda}} 
    - \underset{-H^{\perp}_\mathrm{int, e}}{\underbrace{\sum_{i\in\mathrm{e}} \frac{q_i}{m_i} \mathbf{p}_i\cdot \sum_\lambda A_\lambda \mathbf{f}_\lambda(\mathbf{r}_i) }}\\
    &\,
    \label{eq.Hamiltonian_multipolarcoupling}
    + \underset{H_\mathrm{PSE,c}}{\underbrace{\sum_{\alpha\alpha'}\left(\sum_\lambda\frac{C_{\alpha\lambda}C_{\alpha'\lambda}}{2\varepsilon_0} \right) X_\alpha X_{\alpha'}}} 
    + \underset{H_\mathrm{diam,e}}{\underbrace{\sum_i \frac{q_i^2}{2m_i} \left(\sum_\lambda A_\lambda \mathbf{f}_\lambda(\mathbf{r}_i)\right)^2}}.
\end{align}

In order to perform the diagonalization, let us focus on \(H_\mathrm{c-p} = H_\mathrm{c} + H_\mathrm{PSE,c} + H_\mathrm{f} + H_\mathrm{int, c}^{\perp}\). As can be readily seen from \autoref{eq.Hamiltonian_multipolarcoupling}, \(H_\mathrm{c-p}\) consists of photonic and polarization harmonic oscillators interacting with each other through a momentum-coordinate coupling. For practical reasons, let us perform yet another canonical transformation of the field variables:
\begin{equation}
    \tilde{X}_\lambda = \frac{-\Pi'_\lambda}{\sqrt{\varepsilon_0}\omega_\lambda} \hspace{0.3cm}\text{and}\hspace{0.3cm} \tilde{P}_\lambda = \sqrt{\varepsilon_0}\omega_\lambda A_\lambda.
\end{equation}
With these new variables, \(H_\mathrm{c-p}\) becomes
\begin{align}
    H_\mathrm{c-p} =
    \sum_{\alpha\alpha'}\left[\frac{\delta_{\alpha\alpha'}}{2} P'_\alpha P'_{\alpha'} + \frac{\omega_\alpha^2\delta_{\alpha\alpha'} + \tilde{c}_{\alpha\alpha'}}{2}X_\alpha X_{\alpha'}\right] 
    + \sum_{\lambda\lambda'}\left[\frac{\delta_{\lambda\lambda'}}{2} \tilde{P}_\lambda \tilde{P}_{\lambda'} + \frac{\omega_\lambda^2\delta_{\lambda\lambda'}}{2} \tilde{X}_\lambda \tilde{X}_{\lambda'}\right] + \sum_{\alpha\lambda}\frac{\omega_\lambda C_{\alpha\lambda}}{2\sqrt{\varepsilon_0}} X_\alpha \tilde{X}_\lambda,
\end{align}
where
\begin{equation}
    \tilde{c}_{\alpha\alpha'} = c_{\alpha\alpha'} + \sum_\lambda \frac{C_{\alpha\lambda} C_{\alpha'\lambda}}{\varepsilon_0}.
\end{equation}
Thanks to the last transformation, the photon and cavity oscillators are now coupled through their coordinates. Thus, finding the eigenmodes is reduced to diagonalizing a symmetric quadratic form \(K\), given by
\begin{equation}
    K_{\alpha\alpha'} = \frac{\omega_{\alpha}^2}{2}\delta_{\alpha\alpha'} + \frac{\tilde{c}_{\alpha\alpha'}}{2}, \hspace{0.3cm} K_{\lambda\lambda'} = \frac{\omega_{\lambda}^2}{2}\delta_{\lambda\lambda'}\hspace{0.3cm}\text{and} \hspace{0.3cm} K_{\alpha\lambda} = K_{\lambda\alpha} = \frac{\omega_\lambda }{4\sqrt{\varepsilon_0}}C_{\alpha\lambda}.
\end{equation}
Such a quadratic form can be diagonalized with an orthogonal transformation \(O\), such that \(O^{-1} = O^\mathrm{t}\). Then, \(OKO^\mathrm{t} = D\), where \(D\) is a diagonal matrix with the eigenfrequencies \(\Omega_\eta\) squared. The polaritonic coordinates and momenta are
\begin{subequations}
    \begin{align}
        \beta_\eta &= \sqrt{\frac{\hbar}{2\Omega_\eta}} \left(b^\dagger_\eta + b_\eta\right) = \sum_\alpha O_{\eta\alpha} X_\alpha + \sum_\lambda O_{\eta\lambda} \frac{-\Pi'_\lambda}{\sqrt{\varepsilon_0}\omega_\lambda} \hspace{0.3cm}\text{and}\\
        \xi_\eta &= i\sqrt{\frac{\hbar\Omega_\eta}{2}}\left(b^\dagger_\eta - b_\eta\right) = \sum_\alpha O_{\eta\alpha} P_\alpha + \sum_\lambda O_{\eta\lambda} \sqrt{\varepsilon_0}\omega_\lambda A'_\lambda.
    \end{align}
\end{subequations}
Because the transformation is orthogonal, the inverse relations are
\begin{subequations}
    \begin{align}
        P_\alpha = \sum_\eta O_{\eta\alpha} \xi_\eta, \hspace{0.3cm} \Pi'_\lambda = -\sqrt{\varepsilon_0}\omega_\lambda \sum_\eta O_{\eta\lambda} \beta_\eta, \hspace{0.3cm} 
        X_\alpha = \sum_\eta O_{\eta\alpha} \beta_\eta \hspace{0.3cm}\text{and}\hspace{0.3cm} A_\lambda = \frac{1}{\sqrt{\varepsilon_0}\omega_\lambda}\sum_\eta O_{\eta\lambda} \xi_\eta.
    \end{align}
\end{subequations}
With the above expression of \(A_\lambda\) in terms of the polaritonic momenta \(\xi_\eta\), we can go back to \(H_\mathrm{int, e}^{\perp}\) and write it in the long-wavelength approximation as
\begin{align}
    \nonumber
    H_\mathrm{int, e}^{\perp} &=\,
    \sum_{i\in\mathrm{e}} \frac{q_i}{m_i} \mathbf{p}_i\cdot \sum_{\lambda\eta}\frac{1}{\sqrt{\varepsilon_0}\omega_\lambda} \xi_\eta O_{\eta\lambda} \mathbf{f}_\lambda(\mathbf{r}_\mathrm{e}) = i\sum_{t\eta} \sigma_t \frac{\omega_t}{\Omega_\eta} \left[\sum_\lambda \frac{\Omega_\eta}{\sqrt{\varepsilon_0}\omega_\lambda} O_{\lambda\eta} \mathbf{d}_t \cdot \mathbf{f}_\lambda(\mathbf{r}_\mathrm{e})\right] \xi_\eta + \mathrm{H.c.} \\
    &= -\hbar \sum_{t\eta} \sigma_t \frac{\omega_t}{\Omega_\eta} g_{t\eta}^\perp \left(b_\eta^\dagger - b_\eta\right) + \mathrm{H.c.}.
\end{align}
Here, we have collected the sum over \(\lambda\) in the transverse coupling strength:
\begin{equation}
    g_{t\eta}^{\perp} = \sqrt{\frac{\Omega_\eta}{2\hbar\varepsilon_0}}\sum_\lambda \frac{\Omega_\eta}{\omega_\lambda}  O_{\eta\lambda}\mathbf{d}_t\cdot \mathbf{f}_\lambda(\mathbf{r}_\mathrm{e}),
\end{equation}
From \(g_t^\perp\), we can straightforwardly find the transverse spectral density:
\begin{equation}
    J_t^\perp(\omega) = \sum_\eta \left(g_{t\eta}^{\perp}\right)^2 \delta(\omega-\Omega_\eta) = \frac{1}{2\hbar\varepsilon_0} \sum_\eta \Omega_\eta^3 \sum_{\lambda\lambda'} \frac{ O_{\eta\lambda}  O_{\eta\lambda'}}{\omega_\lambda\omega_{\lambda'}}\left[\mathbf{d}_t\cdot\mathbf{f}_\lambda(\mathbf{r}_\mathrm{e})\right] \left[\mathbf{d}_t\cdot\mathbf{f}_{\lambda'}(\mathbf{r}_\mathrm{e}) \right]\delta(\omega-\Omega_\eta)
\end{equation}
Up to this point, we have successfully written down expressions for \(J_{t, 0}^{\perp}\) and \(J_t^\perp\). The goal is to prove that 
\begin{equation}
    \int_0^\infty \mathrm{d}\omega\ \frac{J_t^\perp(\omega) - J_{t, 0}^{\perp}(\omega)}{J_{t, 0}^{\perp}(\omega)} = 0.
\end{equation}
To that end, let us recall that \(\sum_\eta O_{\eta\lambda}O_{\eta\lambda'}=\delta_{\lambda\lambda'}\), because the transformation is orthogonal. It then follows that 
\begin{align}
    \nonumber
    \int_0^\infty \mathrm{d}\omega \frac{J_t^\perp(\omega)}{\omega^3} 
    &= \int_0^\infty \mathrm{d}\omega \frac{1}{2\hbar\varepsilon_0} \sum_\eta \frac{\Omega_\eta^3}{\omega^3} \sum_{\lambda\lambda'} \frac{O_{\eta\lambda}  O_{\eta\lambda'}}{\omega_\lambda\omega_{\lambda'}}\left[\mathbf{d}_t\cdot\mathbf{f}_\lambda(\mathbf{r}_\mathrm{e})\right] \left[\mathbf{d}_t\cdot\mathbf{f}_{\lambda'}(\mathbf{r}_\mathrm{e}) \right]\delta(\omega-\Omega_\eta) \\
    \nonumber
    &= \frac{1}{2\hbar\varepsilon_0}\sum_{\lambda\lambda'} 
    \left(\sum_\eta O_{\eta\lambda}  O_{\eta\lambda'}\right) \frac{\left[\mathbf{d}_t\cdot\mathbf{f}_\lambda(\mathbf{r}_\mathrm{e})\right] \left[\mathbf{d}_t\cdot\mathbf{f}_{\lambda'}(\mathbf{r}_\mathrm{e}) \right]}{\omega_\lambda\omega_{\lambda'}} = \frac{1}{2\hbar\varepsilon_0} \sum_\lambda \frac{\left[\mathbf{d}_t\cdot\mathbf{f}_\lambda(\mathbf{r}_\mathrm{e})\right]^2}{\omega_\lambda^2} \\
    &= \sum_\lambda \frac{\omega_\lambda}{2\hbar\varepsilon_0} \frac{\left[\mathbf{d}_t\cdot\mathbf{f}_\lambda(\mathbf{r}_\mathrm{e})\right]^2}{\omega_\lambda^3} = \int_0^\infty\mathrm{d}\omega\ \frac{1}{\omega^3} \sum_\lambda \frac{\omega_\lambda}{2\hbar\varepsilon_0} \left[\mathbf{d}_t\cdot\mathbf{f}_\lambda(\mathbf{r}_\mathrm{e})\right]^2 \delta(\omega-\omega_\lambda) = \int_0^\infty \mathrm{d}\omega\ \frac{J_{t, 0}^{\perp}(\omega)}{\omega^3}.
\end{align}
Since \(J_{t, 0}^{\perp}(\omega)\propto\omega^3\), by adding the appropriate factors and rearranging the final equality, we successfully reach the sum rule.

\subsection*{Expansion of the PSE in emitter-mode couplings in the long-wavelength approximation}

We now derive Eq.~(14) of the main text with manipulations similar to the ones above. First, we observe that the PSE only appears when the emitter-photon interaction is written in the multipolar coupling picture. Hence, we take \autoref{eq.Hamiltonian_minimalcoupling} and perform a unitary transformation
\begin{equation}
    U = \exp \left\{-\frac{i}{\hbar}\int\mathrm{d}^3r\ \mathbf{P}_\mathrm{e}^\perp (\mathbf{r})\cdot \mathbf{A}^\perp(\mathbf{r})\right\}
\end{equation}
where 
\begin{equation}
    \mathbf{P}_\mathrm{e}(\mathbf{r}) = \sum_{i\in\mathrm{e}} q_i \mathbf{r}_i \int_0^1\mathrm{d}\sigma\ \delta(\mathbf{r}-\sigma\mathbf{r}_i).
\end{equation}
As a result, the Hamiltonian becomes
\begin{align}
    \nonumber
    H =&\, 
    \underset{H_\mathrm{e}}{\underbrace{\sum_{i\in\mathrm{e}} \frac{\mathbf{p}_i^2}{2m_i} + \sum_{i>j\in\mathrm{e}} \frac{q_i q_j}{4\pi\varepsilon_0 \left|\mathbf{r}_i - \mathbf{r}_j\right|}}} 
    + \underset{H_\mathrm{c}}{\underbrace{\sum_{\alpha\alpha'} \left[\frac{\delta_{\alpha\alpha'}}{2}P_\alpha P_{\alpha'} + \frac{\omega_\alpha^2 \delta_{\alpha\alpha'} + c_{\alpha\alpha'}}{2} X_\alpha X_{\alpha'}\right]}} 
    + \underset{H_\mathrm{int,e}^\parallel}{\underbrace{\sum_\alpha\int\mathrm{d}^3r\ \mathbf{P}_\mathrm{e}(\mathbf{r}) \cdot \bE^\parallel_\alpha(\mathbf{r}) X_\alpha }}\\
    \nonumber
    &\, 
    + \underset{H_\mathrm{f}}{\underbrace{\sum_{\lambda}\left[\frac{1}{2\varepsilon_0}\Pi_\lambda^2 + \frac{\varepsilon_0 \omega_\lambda^2 }{2} A^2_\lambda\right]}} 
    + \underset{H^{\perp, \mathrm{s}}_\mathrm{int}}{\underbrace{\sum_{\alpha\lambda} C_{\alpha\lambda}P_\alpha A_\lambda}} 
    + \underset{H^{\perp}_\mathrm{int, e}}{\underbrace{\frac{1}{\varepsilon_0}\int\mathrm{d}^3r\ \mathbf{P}_\mathrm{e}(\mathbf{r}) \cdot \sum_\lambda \Pi_\lambda \mathbf{f}_\lambda(\mathbf{r})}}\\
    &\,
    \label{eq.Hamiltonian_multipminimcoupling}
    + \underset{H_\mathrm{PSE,e}}{\underbrace{\frac{1}{2\varepsilon_0}\int\mathrm{d}^3r\ \left(\mathbf{P}_\mathrm{e}^\perp(\mathbf{r})\right)^2}} 
    + \underset{H_\mathrm{diam, c}}{\underbrace{\sum_i \frac{q_i^2}{2m_i} \left(\sum_\lambda A_\lambda \mathbf{f}_\lambda(\mathbf{r}_i)\right)^2}},
\end{align}
where the field's canonical momentum is \(\boldsymbol{\Pi}^\perp(\mathbf{r}) = -\left(\varepsilon_0\mathbf{E}^\perp(\mathbf{r})+ \mathbf{P}^\perp_\mathrm{c}(\mathbf{r}) \right)\) and magnetic effects have been neglected. The emitter-field transverse interaction is now given by
\begin{equation}
    \frac{1}{\varepsilon_0} \int\mathrm{d}^3r\ \mathbf{P}_\mathrm{e}^\perp(\mathbf{r})\cdot\sum_\lambda \Pi_\lambda \mathbf{f}_\lambda(\mathbf{r}) \simeq i\hbar \sum_{t\lambda} \sigma_t g_{t\lambda, 0}^{\perp}\left(a_\lambda^\dagger - a_\lambda\right) + \mathrm{H.c.},
\end{equation}
where we have defined
\begin{equation}
    \label{eq.multipolar_g0}
    g_{t\lambda, 0}^{\perp} = \sqrt{\frac{\omega_\lambda}{2\hbar\varepsilon_0}}\mathbf{d}_t\cdot\mathbf{f}_\lambda(\mathbf{r}_\mathrm{e}).
\end{equation}
Note that the above expression corresponds to the interaction strength of the emitter to the ``multipolar coupling photons'', rather than to the ``minimal coupling photons'' in \autoref{eq.minimal_g0}. In spite of this conceptual difference, \autoref{eq.minimal_g0} and \autoref{eq.multipolar_g0} turn out to be the same, as a result of leaving the factor \(\omega_t/\omega_\lambda\) out in the minimal coupling case. A diagonalization procedure of the cavity-photon subsystem similar to the one above can be carried out here again, which yields
\begin{equation}
    \Pi_\lambda = \sqrt{\varepsilon_0}\sum_\eta O'_{\eta\lambda} \xi_\eta,
\end{equation}
and \(O'\) is the corresponding orthogonal transformation. Thus,
\begin{equation}
    \frac{1}{\varepsilon_0} \int\mathrm{d}^3r\ \mathbf{P}_\mathrm{e}^\perp(\mathbf{r})\cdot\sum_\lambda \Pi_\lambda \mathbf{f}_\lambda(\mathbf{r}) \simeq i\hbar \sum_{t\eta} \sigma_t g_{t\eta}^{\perp}\left(b_\eta^\dagger - b_\eta\right) +\mathrm{H.c.}
\end{equation}
Here, the coupling strength is defined as
\begin{equation}
    g_{t\eta}^{\perp} = \sqrt{\frac{\omega_\eta}{2\hbar\varepsilon_0}}\left[\sum_\lambda O_{\eta\lambda}'\mathbf{d}_{t}\cdot \mathbf{f}_\lambda(\mathbf{r}_\mathrm{e})\right].
\end{equation}

Now that both coupling strengths are defined, we expand \(\mathbf{P}_\mathrm{e}^\perp(\mathbf{r})\) in the transverse mode functions (in the long-wavelength approximation):
\begin{equation}
    \mathbf{P}_\mathrm{e}^\perp(\mathbf{r}) \simeq \sum_\lambda \left[\mathbf{d}\cdot \mathbf{f}_\lambda(\mathbf{r}_\mathrm{e})\right] \mathbf{f}_\lambda(\mathbf{r}) = \sum_\lambda \sqrt{\frac{2\hbar\varepsilon_0}{\omega_\lambda}} \sum_t \left(g_{t\lambda, 0}^{\perp} \sigma_t + \mathrm{H.c.}\right) \mathbf{f}_\lambda(\mathbf{r}).
\end{equation}
Thus, the PSE becomes
\begin{align}
    \nonumber
    \frac{1}{2\varepsilon_0} \int\mathrm{d}^3r\ \left(\mathbf{P}_\mathrm{e}^\perp(\mathbf{r})\right)^2 
    \simeq&\, 
    \hbar\sum_\lambda \frac{\left(\sum_t g_{t\lambda,0}^{\perp} \sigma_t + \mathrm{H.c.}\right)^2}{\omega_\lambda} =
    \hbar\sum_{tt'} \sum_\lambda \frac{g_{t\lambda,0}^{\perp} g_{t'\lambda,0}^{\perp} \sigma_t\sigma_{t'} + g_{t\lambda,0}^{\perp} \left(g_{t'\lambda,0}^{\perp}\right)^* \sigma_t\sigma_{t'}^\dagger + \mathrm{H.c.} }{\omega_\lambda}\\
    \nonumber
    =&\, 
    \hbar\sum_{tt'} \sum_{\lambda\lambda'} \frac{g_{t\lambda,0}^{\perp} g_{t'\lambda',0}^{\perp} \sigma_t\sigma_{t'} + g_{t\lambda,0}^{\perp} \left(g_{t'\lambda',0}^{\perp}\right)^* \sigma_t\sigma_{t'}^\dagger + \mathrm{H.c.} }{\omega_\lambda} \sum_{\eta}O'_{\eta\lambda}O'_{\eta\lambda'}\\
    =&\, 
    \hbar\sum_{tt'} \sum_{\eta} \frac{g_{t\eta}^{\perp} g_{t'\eta}^{\perp} \sigma_t\sigma_{t'} + g_{t\eta}^{\perp} \left(g_{t'\eta}^{\perp}\right)^* \sigma_t\sigma_{t'}^\dagger + \mathrm{H.c.} }{\omega_\eta} = \hbar\sum_\eta \frac{\left(\sum_t g_{t\eta}^\perp \sigma_t + \mathrm{H.c.}\right)^2}{\omega_\eta}.
\end{align}
Here, we have used that \(\sum_{\eta}O'_{\eta\lambda}O'_{\eta\lambda'}=\delta_{\lambda\lambda'}\), and that \(g_{t\eta}^{\perp} = \sum_\lambda \sqrt{\frac{\omega_\eta}{\omega_\lambda}} O_{\eta\lambda} g_{t\lambda,0}^{\perp}\). The final expression is precisely Eq.~(14) of the main text.

\section*{Coulomb interaction between the charged spheres}

\begin{figure}[htbp]
    \centering
    \includegraphics[scale=1]{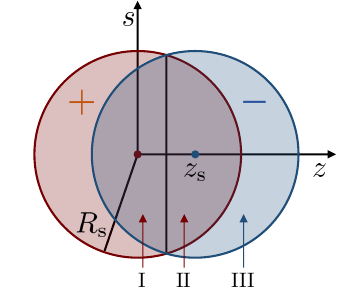}
    \caption{Graphical representation of the integration regions.}
    \label{fig.regions}
\end{figure}

In essence, we have to integrate the electrostatic potential due to the ionic sphere \(\phi_+(\mathbf{r})\) [see Eq.~(18) of the main text] over the volume of the electronic sphere. The complication arises from the off-set of their centers.  We begin by splitting the electronic sphere's volume in three parts, defined by \(z_\mathrm{s}-R_\mathrm{s} \leq z \leq z_\mathrm{s}/2\) (I), \(z_\mathrm{s}/2<z\leq R_\mathrm{s}\) (II) and the non-overlapping part (III) (see \autoref{fig.regions}). 

For regions I and II, we use cylindrical coordinates where the angular integration yields a \(2\pi\) factor. Then, we have 
\begin{align}
    \nonumber
    \int_\mathrm{I/II}\mathrm{d}^3r\ \rho(\mathbf{r})\phi_+(\mathbf{r}) 
    &= \frac{\pi\rho^2}{3\varepsilon_0} \int_\mathrm{I/II}\mathrm{d}z\ \int_0^{S_\mathrm{I/II}(z)}\mathrm{d}s\ s\left(3R_\mathrm{s}^2-s^2-z^2\right) \\
    &= \frac{\pi \rho^2}{12\varepsilon_0}\int_\mathrm{I/II}\mathrm{d}z\ \left[\left(6R_\mathrm{s}^2 - 2z^2\right)S_\mathrm{I/II}^2(z) - S_\mathrm{I/II}^4(z)\right],
\end{align}
where \(S_\mathrm{I}(z) = \sqrt{R_\mathrm{s}^2 + (z-z_\mathrm{s})^2}\) and \(S_\mathrm{II}(z) = \sqrt{R_\mathrm{s}^2 + z^2}\). The integrands over \(z\) of regions I and II are then simply polynomic functions. After lengthy manipulations, we obtain
\begin{equation}
    \int_\mathrm{I+II}\mathrm{d}^3r\ \rho(\mathbf{r})\phi_+(\mathbf{r}) = \frac{\pi\rho^2}{3\varepsilon_0}\left[\frac{8R_\mathrm{s}^5}{5} - R_\mathrm{s}^4 z_\mathrm{s} - \frac{R_\mathrm{s}^3 z^2_\mathrm{s}}{3} + \frac{R_\mathrm{s}^2 z_\mathrm{s}^3}{4} - \frac{z_\mathrm{s}^5}{120}\right].
\end{equation}
For region III, we use spherical coordinates. The azimuthal angle integral again provides a \(2\pi\) factor. The integration limits for the polar angle \(\vartheta\) are \(0\) and \(\vartheta_0\), where \(\cos\vartheta_0=\frac{z_\mathrm{s}}{2R_\mathrm{s}}\), and the radial integration limits are \(R_\mathrm{s}\) and \(D(\cos\vartheta) = z_\mathrm{s}\cos\vartheta + \sqrt{z_\mathrm{s}^2\cos^2\vartheta + R_\mathrm{s}^2 - z_\mathrm{s}^2}\). Thus,
\begin{align}
    \int_\mathrm{III}\mathrm{d}^3r\ \rho(\mathbf{r})\phi_+(\mathbf{r}) &= \frac{2\pi\rho^2 R_\mathrm{s}^3}{3\varepsilon_0} \int_{0}^{\vartheta_0}\mathrm{d}\vartheta \int_{R_\mathrm{s}}^{D(\cos\vartheta)}\mathrm{d}r\ r^2 \frac{1}{r}
    = \frac{\pi\rho^2R_\mathrm{s}^3}{3\varepsilon_0}\int_{\frac{z_\mathrm{s}}{2R_\mathrm{s}}}^{1}\mathrm{d}x\ \left[D^2(x) - R_\mathrm{s}^2\right] = \frac{\pi\rho^2}{3\varepsilon_0}\left[R_\mathrm{s}^4z_\mathrm{s} - \frac{R^3_\mathrm{s}z_\mathrm{s}^2}{3}\right]
\end{align}
where \(x=\cos\vartheta\), and the intermediate steps are immediate integrals. Then, the whole electrostatic interaction between the spheres is
\begin{equation}
    \int\mathrm{d}^3r\ \rho(\mathbf{r})\phi_+(\mathbf{r}) = \frac{\pi\rho^2}{3\varepsilon_0} \left[\frac{8R_\mathrm{s}^5}{5} - \frac{2R_\mathrm{s}^3z_\mathrm{s}^2}{3} + \frac{R_\mathrm{s}^2|z_\mathrm{s}|^3}{4} - \frac{|z_\mathrm{s}|^5}{120}\right].
\end{equation}

In the small oscillation limit, we retain only the constant and quadratic terms. The constant cancels the electrostatic self-energies of the spheres, and the quadratic term gives rise to a harmonic restoring force. Expressed in terms of the electronic sphere's mass and the plasma frequency \(\Omega_\mathrm{p}=\sqrt{\frac{\rho e}{m_\mathrm{e}\varepsilon_0}}\), we recover 
\begin{equation}
    \int\mathrm{d}^3r\ \rho(\mathbf{r})\phi_+(\mathbf{r}) = \frac{8\pi\rho^2R_\mathrm{s}^5}{15\varepsilon_0} - \frac{m_\mathrm{s}}{2}\frac{\Omega_\mathrm{p}^2}{3}z_\mathrm{s}^2.
\end{equation}

\section*{Cavity-photon interaction}

The cavity-photon interaction is given by the generalization of \(\mathbf{p}_i - q_i \mathbf{A}^\perp(\mathbf{r}_i)\) to an extended charge distribution that appears in \(H_\mathrm{s}\), in Eq.~(17). We expand the vector potential in the spherical-wave basis and find Eq.~(24). Here, we perform the calculation of \(I_{lm}^{(\lambda)}(\omega)\) from the main text, which is essentially the integral of the \(z\) component of the basis functions over the volume of the negatively charged sphere, which is slightly displaced along the \(z\) axis. As we will see in this section, we can neglect the displacement and simply use \(\rho(\mathbf{r}) = \rho\theta(|\mathbf{r}|-R_\mathrm{s})\). The price of this approximation is that terms \(\mathcal{O}(z_\mathrm{s})\) and beyond are discarded, which is fine if the constant term is non-zero and the oscillations are small.

With the above considerations, let us evaluate the \(\lambda=\mathrm{tm}\) integral first:
\begin{align}
    \nonumber
    I_{lm}^{(\mathrm{tm})}(\omega) 
    &= \int\mathrm{d}^3r\ \rho(\mathbf{r})\hat{\mathbf{z}}\cdot\mathbf{f}_{lm}^{(\mathrm{tm})}(\omega, \mathbf{r}) = \frac{c}{\omega}\int\mathrm{d}^3r\ \rho(\mathbf{r})\hat{\mathbf{z}}\cdot\nabla\times\mathbf{f}_{lm}^{(\mathrm{te})}(\omega, \mathbf{r})\\
    \nonumber
    &=\frac{c}{\omega}\int\mathrm{d}^3r\ \left[-\nabla\cdot\left(\rho(\mathbf{r})\hat{\mathbf{z}}\times \mathbf{f}_{lm}^{(\mathrm{te})}(\omega, \mathbf{r})\right) + \mathbf{f}_{lm}^{(\mathrm{te})}(\omega, \mathbf{r}) \cdot\nabla\times\Big(\rho(\mathbf{r})\hat{\mathbf{z}}\Big)\right] = \frac{c}{\omega}\int\mathrm{d}^3r\  \mathbf{f}_{lm}^{(\mathrm{te})}(\omega, \mathbf{r}) \cdot\Big(\nabla\rho(\mathbf{r})\Big)\times\hat{\mathbf{z}} \\
    \nonumber
    &= -\frac{\rho c}{\omega}\int_0^{2\pi}\mathrm{d}\varphi\int_0^\pi\mathrm{d}\vartheta\ \sin\vartheta \int_0^\infty \mathrm{d}r\ r^2\delta(r-R_\mathrm{s}) \mathbf{f}_{lm}^{(\mathrm{te})}(\omega, \mathbf{r}) \cdot\hat{\mathbf{r}}\times\left(\hat{\mathbf{r}}\cos\vartheta - \hat{\boldsymbol{\vartheta}}\sin\vartheta\right) \\
    \nonumber
    &=  \frac{\sqrt{2}\rho}{\sqrt{l(l+1)\pi c}}\int_0^{2\pi}\mathrm{d}\varphi\int_0^\pi\mathrm{d}\vartheta\ \sin^2\vartheta \int_0^\infty \mathrm{d}r\ r^2\delta(r-R_\mathrm{s}) j_l\left(\frac{\omega |\mathbf{r}|}{c}\right) \left[\hat{\boldsymbol{\vartheta}}\frac{im}{\sin\vartheta}-\hat{\boldsymbol{\varphi}}\partial_\vartheta\right] Y_l^m(\vartheta, \varphi) \cdot \hat{\boldsymbol{\varphi}}\\
    \nonumber
    &= -\frac{\sqrt{2}\rho R_\mathrm{s}^2j_l\left(\frac{\omega R_\mathrm{s}}{c}\right)}{\sqrt{l(l+1)\pi c}}\int_0^{2\pi}\mathrm{d}\varphi\int_0^\pi\mathrm{d}\vartheta\ \sin^2\vartheta\ \partial_\vartheta Y_l^m(\vartheta, \varphi) = \frac{\sqrt{8}\rho R_\mathrm{s}^2j_l\left(\frac{\omega R_\mathrm{s}}{c}\right)}{\sqrt{l(l+1)\pi c}}\int\mathrm{d}\Omega\ \cos\vartheta\ Y_l^m(\vartheta, \varphi) \\
    \nonumber
    &= \frac{4\sqrt{2}\rho R_\mathrm{s}^2j_l\left(\frac{\omega R_\mathrm{s}}{c}\right)}{\sqrt{3l(l+1) c}}\int\mathrm{d}\Omega\ Y_1^0(\vartheta, \varphi) Y_l^m(\vartheta, \varphi) = \frac{4\rho R_\mathrm{s}^2}{\sqrt{3 c}}j_1\left(\frac{\omega R_\mathrm{s}}{c}\right)\delta_{l1}\delta_{m0}.
\end{align}
To arrive at this expression, we have first used the relation between tm and te mode functions, to take advantage of the simpler expression of the te ones. Then, we have integrated the curl by parts. The divergence in the second line vanishes because \(\rho=0\) at \(|\mathbf{r}|>R_\mathrm{s}\). The curl in the second line can be rewritten in terms of the gradient of the charge density, which yields \(-\hat{\mathbf{r}}\delta(r-R_\mathrm{s})\). We next express \(\hat{\mathbf{z}}\) in spherical coordinates through \(\hat{\mathbf{z}}=\hat{\mathbf{r}}\cos\vartheta - \hat{\boldsymbol{\vartheta}}\sin\vartheta\), and take the vector product \(\hat{\mathbf{r}}\times\hat{\boldsymbol{\vartheta}}=\hat{\boldsymbol{\varphi}}\). Recalling the definition of \(\mathbf{f}^\mathrm{(te)}_{lm}\) from Eq.~(23a) of the main text, we trivially integrate over \(r\). The remaining angular integral can be evaluated through integration by parts, and then using the orthonormality of the spherical harmonics. Thus, only the \(l=1, m=0\) element is non-zero. Strictly speaking, as mentioned above, there are further contributions of \(\mathcal{O}(z_\mathrm{s})\), but those are negligible in the small oscillation limit.

Let us finish this section by evaluating the \(\lambda=\mathrm{te}\) terms:
\begin{align}
    \nonumber
    I_{lm}^{(\mathrm{te})}(\omega) 
    &= \int\mathrm{d}^3r\ \rho(\mathbf{r})\hat{\mathbf{z}}\cdot\mathbf{f}_{lm}^{(\mathrm{te})}(\omega, \mathbf{r}) = \int\mathrm{d}^3r\ \rho(\mathbf{r}) \frac{\omega}{c}\sqrt{\frac{2}{l(l+1)\pi c}} j_l\left(\frac{\omega |\mathbf{r}|}{c}\right) \hat{\mathbf{z}}\cdot \left[\hat{\boldsymbol{\vartheta}}\frac{im}{\sin\vartheta}-\hat{\boldsymbol{\varphi}}\partial_\vartheta\right] Y_l^m(\vartheta, \varphi) \\
    \nonumber
    &= \frac{-im\rho\omega\sqrt{2}}{\sqrt{l(l+1)\pi c^3}} \int\mathrm{d}\Omega\ Y_{l}^m(\vartheta,\varphi) \int_0^{R_\mathrm{s}}\mathrm{d}r\ r^2 j_l\left(\frac{\omega r}{c}\right) = 0,
\end{align}
which vanishes because of the angular integral. Consequently, in the limit of small oscillations, the sphere only couples to modes with \(\lambda=\mathrm{tm}\), \(l=1\) and \(m=0\).

\section*{Integrals involving spherical Bessel functions}

We analytically calculate here certain quantities that are necessary for the simple analytical example of the main text. To solve the integrals involved, we make heavy use of the properties of Fourier transforms. Thus, we begin by stating the basic definitions and solving several useful integrals. Then, we proceed to calculate the relevant quantities for the model.

\subsection*{Fourier relations and basic integrals}

First, the Fourier transform is defined as
\begin{equation}
    \mathcal{F}[f(x)](k) = \frac{1}{\sqrt{2\pi}} \int_{-\infty}^{\infty}\mathrm{d}x\ f(x)e^{-ikx}.
\end{equation}
We will require two basic properties, namely,
\begin{align}
    \mathcal{F}[f(x-x_0)](k) &= e^{-ikx_0}\mathcal{F}[f](k) \\
    \mathcal{F}[f(ax)](k) &= \frac{1}{|a|}\mathcal{F}[f]\left(\frac{k}{a}\right) .
\end{align}
With the above definition of the Fourier transform, the convolution theorem is expressed as 
\begin{equation}
    \mathcal{F}[f(x)g(x)](k) = \frac{1}{\sqrt{2\pi}} \left\{\mathcal{F}[f(x)]*\mathcal{F}[g(x)]\right\}(k).
\end{equation}
Next, as seen in the previous section, the spherical Bessel function of the first kind \(j_1(x) = \frac{\sin x}{x^2}-\frac{\cos x}{x}\) plays an important role in the integrals that have to be calculated. Therefore, we add its Fourier transform to the list of relations:
\begin{equation}
    \mathcal{F}[j_1(x)](k) = \begin{cases}
        i\sqrt{\frac{\pi}{2}}k & \text{if } -1<k<1\\
        0 & \text{otherwise.}
    \end{cases}
\end{equation}
The last important transforms that will play a role are
\begin{align}
    \mathcal{F}\left[\frac{1}{x}\right](k) &= -i\sqrt{\frac{\pi}{2}}\mathrm{sign}(k)
\end{align}

With the above basic properties, we now calculate three Fourier transforms that will determine the result of all the integrals to come. 
\begin{itemize}
    \item The first one is 
    \begin{align}
        \nonumber
        \mathcal{F}\left[\frac{1}{x^2-x_0^2}\right](k) &= \frac{1}{2|x_0|}\mathcal{F}\left[\frac{1}{x-x_0} - \frac{1}{x+x_0}\right](k) = \frac{e^{-ikx_0} - e^{ikx_0}}{2|x_0|}\mathcal{F}\left[\frac{1}{x}\right](k) \\
        &= -\sqrt{\frac{\pi}{2}}\frac{\sin(|k|x_0)}{|x_0|}.
    \end{align}
    \item The second transform is \(\mathcal{F}[j_1(x)j_1(yx)](k)\), with \(y\geq1\), whose value can be found with the convolution theorem for \(k\geq 0\):
    \begin{align*}
        \nonumber
        \mathcal{F}[j_1(x)j_1(yx)](k) &= \frac{1}{\sqrt{2\pi}} \int_{-\infty}^{\infty}\mathrm{d}k'\  \mathcal{F}[j_1(x)](k')\  \mathcal{F}[j_1(yx)](k-k') \\
        \nonumber
        &= \sqrt{\frac{\pi}{2}}\frac{1}{2y^2}
        \begin{cases}
            \displaystyle
            \int_{-1}^{1} \mathrm{d}k'\ \left({k'}^2 - kk'\right) & \text{if } 0\leq k < y-1 \\
            \displaystyle
            \int_{-1}^{k-y} \mathrm{d}k'\ \left({k'}^2 - kk'\right) & \text{if } y-1\leq k < y+1 \\
            0 & \text{otherwise}
        \end{cases} \\
        &= \sqrt{\frac{\pi}{2}}\frac{1}{12y^2}
        \begin{cases}
            \displaystyle
            4 & \text{if } 0\leq k < y-1 \\
            \displaystyle
            k^3 - 3(y^2+1)k + 2(y^3+1)  & \text{if } y-1\leq k < y+1 \\
            0 & \text{otherwise}.
        \end{cases}
    \end{align*}
    Due to the symmetry properties of the functions involved, we can straightforwardly extend the result for negative \(k\):
    \begin{equation} \label{eq.second_relation}
        \mathcal{F}[j_1(x)j_1(yx)](k) = \sqrt{\frac{\pi}{2}}\frac{1}{12y^2}
        \begin{cases}
            \displaystyle
            4 & \text{if } |k| < y-1 \\
            \displaystyle
            |k|^3 - 3(y^2+1)|k| + 2(y^3+1)  & \text{if } y-1\leq |k| < y+1 \\
            0 & \text{otherwise}.
        \end{cases}
    \end{equation}
    \item The third relation needed is more complicated. Fortunately, the calculation is significantly simplified because only the \(k=0\) value is actually required:
    \begin{align} 
        \nonumber
        \mathcal{F}\left[\frac{j_1(x)j_1(yx)}{x^2-x_0^2}\right](k=0) 
        &= \frac{1}{\sqrt{2\pi}}\int_{-\infty}^{\infty}\mathrm{d}k' \ \mathcal{F}[j_1(x)j_1(yx)](k') \ \mathcal{F}[(x^2-x_0^2)^{-1}](k-k')\Big|_{k=0} \\
        \nonumber
        &= -\frac{2}{\sqrt{2\pi}} \int_{0}^{\infty}\mathrm{d}k' \ \mathcal{F}[j_1(x)j_1(yx)](k') \ \mathcal{F}[(x^2-x_0^2)^{-1}](k-k')\Big|_{k=0} \\
        \nonumber
        &= -\sqrt{\frac{\pi}{2}}\frac{1}{12y^2x_0} \left[\int_{0}^{y-1}\mathrm{d}k'\ 4 \sin(k'x_0) + \int_{y-1}^{y+1}\mathrm{d}k'\ \left({k'}^3-3(y^2-1)k' + 2(y^3+1)\right)\right] \\
        \nonumber
        &= -\sqrt{\frac{\pi}{2}} \frac{1}{x^2_0 y^2}\left[\frac{1}{3} + \frac{(x_0\cos x_0 -\sin x_0)(\cos(yx_0) + yx_0\sin(yx_0))}{x_0^3}\right] \\
        \label{eq.third_relation}
        &= -\sqrt{\frac{\pi}{2}} \frac{1}{x^2_0 y^2}\left[\frac{1}{3} + y^2x_0j_1(x_0)y_1(yx_0)\right],
    \end{align}
    where \(y_1(x)=-\frac{\cos x}{x^2}-\frac{\sin x}{x}\) is the first spherical Bessel function of the second kind. We have used the symmetry of the integrand and several steps of simple, but long, integration by parts, together with the trigonometric angle sum relations. We are now ready to evaluate all the following integrals by suitably applying the above expressions.
\end{itemize}
    
\subsection*{Sphere's polarization self-energy shift}

In the multipolar coupling picture, Eq.~(28) of the main text, the square of the bare frequency of the sphere is renormalized by
\begin{align}
    \nonumber
    \frac{16\rho^2R_\mathrm{s}^4}{3\varepsilon_0m_\mathrm{s}c}\int_0^\infty\mathrm{d}\omega\ j_1^2\left(\frac{\omega R_\mathrm{s}}{c}\right) 
    &= \frac{8\rho^2R_\mathrm{s}^3}{3\varepsilon_0m_\mathrm{s}}\int_{-\infty}^\infty\mathrm{d}x \ j_1^2\left(x\right) = \frac{8\rho^2R_\mathrm{s}^3}{3\varepsilon_0m_\mathrm{s}}\sqrt{2\pi} \mathcal{F}[j_1^2(x)](k=0).
\end{align} 
The Fourier transform can be evaluated by setting \(y=1\) and \(k = 0\) in \autoref{eq.second_relation}. Then, 
\begin{align}
    \frac{16\rho^2R_\mathrm{s}^4}{3\varepsilon_0m_\mathrm{s}c}\int_0^\infty\mathrm{d}\omega\ j_1^2\left(\frac{\omega R_\mathrm{s}}{c}\right) 
    = \frac{8\rho^2 R_\mathrm{s}^3}{3\varepsilon_0m_\mathrm{s}}\frac{\pi}{3} = \frac{2\rho e}{3\varepsilon_0 m_e} = \frac{2\Omega_\mathrm{p}^2}{3}.
\end{align} 

\subsection*{Fano diagonalization energy shift: \(F\)}

The Fano diagonalization procedure involves calculating the integral of Eq.~(34b):
\begin{align}
    \nonumber
    F(\Omega) &= \mathrm{PV}\int_0^\infty \mathrm{d}\omega\ \frac{\gamma^2(\omega)}{\Omega^2-\omega^2} 
    = \frac{2\Omega_\mathrm{p}^2R_\mathrm{s}}{\pi c} \mathrm{PV}\int_{-\infty}^{\infty}\mathrm{d}\omega\ \frac{\omega^2 j^2_1\left(\frac{\omega R_\mathrm{s}}{c}\right)}{\Omega^2-\omega^2} = -\frac{2\Omega_\mathrm{p}^2R_\mathrm{s}}{\pi c} \mathrm{PV}\int_{-\infty}^{\infty}\mathrm{d}\omega\ \left[j^2_1\left(\frac{\omega R_\mathrm{s}}{c}\right) + \frac{\Omega^2 j^2_1\left(\frac{\omega R_\mathrm{s}}{c}\right)}{\omega^2-\Omega^2} \right] \\
    \nonumber
    &= -\frac{2\Omega_\mathrm{p}^2}{\pi}\sqrt{2\pi} \left\{\mathcal{F}[j_1^2(x)](k=0) +\left(\frac{\Omega R_\mathrm{s}}{c}\right)^2 \mathcal{F}\left[\frac{j_1^2(x)}{x^2-\left(\frac{\Omega R_\mathrm{s}}{c}\right)^2}\right](k=0)\right\},
\end{align}
which is easily evaluated with \autoref{eq.second_relation} and \autoref{eq.third_relation}, evaluated at \(x_0 = \frac{\Omega R_\mathrm{s}}{c}\), \(y = 1\) and \(k=0\). Then,
\begin{equation}
    F(\Omega) = \mathrm{PV}\int_0^\infty \mathrm{d}\omega\ \frac{\gamma^2(\omega)}{\Omega^2-\omega^2} = 2\Omega_\mathrm{p}^2 \frac{\Omega R_\mathrm{s}}{c}j_1\left(\frac{\Omega R_\mathrm{s}}{c}\right)y_1\left(\frac{\Omega R_\mathrm{s}}{c}\right).
\end{equation}

\subsection*{Fano diagonalization coefficient: \(c_1\)}

From the commutator \([\beta(\Omega),\xi(\Omega')]=i\hbar\delta(\Omega-\Omega')\) and Eqs.~(33a), (33b) and (34a) of the main text, we can find \(c_1(\Omega)\), Eq.~(35). This procedure is done in the original reference by Fano~\cite{fanoDiagonalization1961}, but the details are slightly more involved here because the diagonal matrix elements are \(\Omega^2\) rather than \(\Omega\). For this reason, we carry it out explicitly. Plugging the main text equations in the commutation relation, we directly arrive at
\begin{align}
    \nonumber
    \frac{\delta(\Omega-\Omega')}{c_1(\Omega)c_1(\Omega')} = 1 + \int\mathrm{d}\omega\ \left[\mathrm{PV}\frac{1}{\Omega-\omega} + k(\Omega)\delta(\Omega-\omega)\right]\left[\mathrm{PV}\frac{1}{\Omega'-\omega} + k(\Omega')\delta(\Omega'-\omega)\right]\gamma^2(\omega),
\end{align}
where we have defined \(k(\Omega) = \frac{\Omega^2-\Omega_\mathrm{p}^2-F(\Omega)}{\gamma^2(\Omega)}\). We expand the product inside the integral and obtain
\begin{align}
    \nonumber
    \frac{\delta(\Omega-\Omega')}{c_1(\Omega)c_1(\Omega')} 
    &= 1 + \frac{k(\Omega')\gamma^2(\Omega')-k(\Omega)\gamma^2(\Omega)}{\Omega^2-{\Omega'}^2} + k^2(\Omega)\gamma^2(\Omega)\delta(\Omega-\Omega') + \mathrm{PV}\int\mathrm{d}\omega\ \frac{\gamma^2(\omega)}{(\Omega^2-\omega^2)({\Omega'}^2-\omega^2)}\\
    \nonumber
    &= \left[k^2(\Omega)+\left(\frac{\pi}{2\Omega}\right)^2\right]\gamma^2(\Omega)\delta(\Omega-\Omega') + \frac{F(\Omega)-F(\Omega')}{\Omega^2-{\Omega'}^2} + \mathrm{PV}\int\mathrm{d}\omega\ \frac{\frac{\gamma^2(\omega)}{\Omega-\Omega'}}{(\Omega+\omega)(\Omega'+\omega)}\left[\frac{1}{\Omega'-\omega}-\frac{1}{\Omega-\omega}\right].
\end{align}
To reach the second line, we have used the following partial fraction decomposition~\cite{fanoDiagonalization1961}:
\begin{equation}
    \mathrm{PV} \frac{1}{(\Omega-\omega)(\Omega'-\omega)} = \frac{1}{\Omega-\Omega'} \mathrm{PV}\left[\frac{1}{\Omega'-\omega} - \frac{1}{\Omega-\omega}\right] + \pi^2 \delta(\Omega'-\omega)\delta(\Omega-\omega).
\end{equation}
Next, we show that the last two terms cancel:
\begin{align*}
    \nonumber
    \mathrm{PV}\int_0^\infty\mathrm{d}\omega\ &\frac{\frac{\gamma^2(\omega)}{\Omega-\Omega'}}{(\Omega'+\omega)(\Omega+\omega)}\left[\frac{1}{\Omega'-\omega}-\frac{1}{\Omega-\omega}\right] = \frac{1}{\Omega^2-{\Omega'}^2}\mathrm{PV}\int_{0}^{\infty}\mathrm{d}\omega\ \frac{\gamma^2(\omega)}{(\Omega'+\omega)(\Omega+\omega)}\left[\frac{\Omega+\Omega'}{\Omega'-\omega}-\frac{\Omega+\Omega'}{\Omega-\omega}\right]\\
    &=\frac{1}{\Omega^2-{\Omega'}^2}\mathrm{PV}\int_{0}^{\infty}\mathrm{d}\omega\ \gamma^2(\omega)\left[\frac{1}{{\Omega'}^2-\omega^2}+\frac{1}{(\Omega'+\omega)(\Omega+\omega)}-\frac{1}{{\Omega}^2-\omega^2}-\frac{1}{(\Omega'+\omega)(\Omega+\omega)}\right] \\
    &= \frac{1}{\Omega^2-{\Omega'}^2}\left[F(\Omega') - F(\Omega)\right].
\end{align*}
Therefore,
\begin{equation}
    c_1^2(\Omega) = \frac{1}{\left[k^2(\Omega) + \left(\frac{\pi}{2\Omega}\right)^2\right]\gamma^2(\Omega)} \implies c_1(\Omega) = \frac{\gamma(\Omega)}{\sqrt{(\Omega^2-\Omega_\mathrm{p}^2-F(\Omega))^2 + \left(\frac{\pi}{2\Omega}\right)^2 \gamma^4(\Omega)}}.
\end{equation}
Note that the last expression carries a sign choice with it. Any other choice would be fine as well, as long as it is consistently kept everywhere.

\subsection*{Emitter-photon interaction}

The interaction between the emitter and the photons, expressed in terms of the polariton eigenmodes, is given by Eq.~(41). The integral there is what we calculate now. In the case considered in the main text, an emitter placed along the \(z\) axis with \(\mathbf{d}_t\parallel \hat{\mathbf{z}}\), we have
\begin{align}
    \nonumber
    g^\perp_t(\Omega) 
    &= |\mathbf{d}_t|\sqrt{\frac{\Omega^3}{2\hbar\varepsilon_0}}\int\mathrm{d}\omega\ \frac{c_2(\Omega, \omega) \hat{\mathbf{z}}\cdot\mathbf{f}_{10}^\mathrm{(tm)}(\omega, \mathbf{r}_\mathrm{e})}{\omega} \\
    \nonumber
    &= c_1(\Omega)\frac{|\mathbf{d}_t|}{|\mathbf{r}_\mathrm{e}|}\sqrt{\frac{3\Omega^3}{2\hbar\pi^2\varepsilon_0c}}\int\mathrm{d}\omega\ \left[\mathrm{PV}\frac{1}{\Omega^2-\omega^2} + \frac{\Omega^2-\Omega_\mathrm{p}^2-F(\Omega)}{\gamma^2(\Omega)}\delta(\Omega-\omega)\right]\frac{\gamma(\omega)}{\omega}
        j_1\left(\frac{\omega |\mathbf{r}_\mathrm{e}|}{c}\right).
\end{align}
Here, we have used that \(\hat{\mathbf{r}}\cdot\mathbf{f}_{lm}^\mathrm{(tm)}(\omega, \mathbf{r}) = \frac{1}{|\mathbf{r}|}\sqrt{\frac{2l(l+1)}{\pi c}}j_l\left(\frac{\omega |\mathbf{r}|}{c}\right) Y_l^m(\vartheta, \varphi)\)~\cite{steckQuantumOptics2007}. The second term in the square brackets yields
\begin{equation}\label{eq.bg}
    \frac{|\mathbf{d}_t|}{|\mathbf{r}_\mathrm{e}|}\sqrt{\frac{3\Omega^3}{2\hbar\pi^2\varepsilon_0c}}  c_1(\Omega)
    \frac{\left(\Omega^2-\Omega_\mathrm{p}^2-F(\Omega)\right)j_1\left(\frac{\Omega |\mathbf{r}_\mathrm{e}|}{c}\right)}{\Omega\gamma(\Omega)},
\end{equation}
and the first is equal to
\begin{align}
    \nonumber
    -\frac{|\mathbf{d}_t|}{|\mathbf{r}_\mathrm{e}|}&\sqrt{\frac{3\Omega^3}{2\hbar\pi^2\varepsilon_0c}} 2\Omega_\mathrm{p} \sqrt{\frac{R_\mathrm{s}}{\pi c}}
    c_1(\Omega)\ \mathrm{PV}\int\mathrm{d}\omega\ \frac{j_1\left(\frac{\omega R_\mathrm{s}}{c}\right)j_1\left(\frac{\omega |\mathbf{r}_\mathrm{e}|}{c}\right)}{\omega^2-\Omega^2} \\
    \nonumber
    &= -\frac{|\mathbf{d}_t|}{|\mathbf{r}_\mathrm{e}|}\sqrt{\frac{3\Omega^3}{2\hbar\pi^2\varepsilon_0c}} \Omega_\mathrm{p} \sqrt{\frac{R^3_\mathrm{s}}{\pi c^3}} 
    \sqrt{2\pi} c_1(\Omega)  \mathcal{F}\left[\frac{j_1(x)j_1\left(\frac{|\mathbf{r}_\mathrm{e}|}{R_\mathrm{s}}x\right)}{x^2-\left(\frac{\Omega R_\mathrm{s}}{c}\right)^2}\right] \\
    &= \frac{|\mathbf{d}_t|}{|\mathbf{r}_\mathrm{e}|}\sqrt{\frac{3\Omega^3}{2\hbar\pi^2\varepsilon_0c}}c_1(\Omega)  
    \frac{\sqrt{\pi cR^3_\mathrm{s}}}{\Omega^2 |\mathbf{r}_\mathrm{e}|^2}\left[\frac{1}{3} + \frac{\Omega|\mathbf{r}_\mathrm{e}|^2 j_1\left(\frac{\Omega R_\mathrm{s}}{c}\right)y_1\left(\frac{\Omega |\mathbf{r}_\mathrm{e}|}{c}\right)}{cR_\mathrm{s}}\right].
\end{align}
Note that, to retrieve the Fourier transform, we extend the integration limits to \(-\infty\) and \(\infty\), with the corresponding cancellation of a factor of 2. Both terms together represent Eq.~(41) of the main text.

\subsection*{Polaritonic polarization self-energy}

The integral in Eq.~(47) has an analytical expression in terms of the sine integral \(\mathrm{Si}(x)\). However, the limit that interests us is \(|\mathbf{r}_\mathrm{e}|\) smaller than \(c/\omega\). In said limit, we can expand \(j_1(x)\simeq \frac{x}{3}\). Then,
\begin{align}
    H_\mathrm{PSE}^\mathrm{expl.} 
    &\simeq \hbar|\hat{\mathbf{z}}\cdot\mathbf{d}|^2 \int_0^{\Omega_\mathrm{p}}\mathrm{d}\omega\ \frac{3j_1^2\left(\frac{\omega|\mathbf{r}_\mathrm{e}|}{c}\right)}{2\hbar\varepsilon_0\pi^2c|\mathbf{r}_\mathrm{e}|^2} 
    \simeq \int_0^{\Omega_\mathrm{p}}\mathrm{d}\omega\ \frac{|\hat{\mathbf{z}}\cdot\mathbf{d}|^2 \omega^2}{6\varepsilon_0\pi^2c^3} 
    \simeq \frac{|\hat{\mathbf{z}}\cdot\mathbf{d}|^2\Omega_\mathrm{p}^3}{18\varepsilon_0\pi^2c^3}
    &\simeq 10^{-5}~ \mathrm{eV}/(e\cdot\mathrm{nm})^2 \times |\hat{\mathbf{z}}\cdot\mathbf{d}|^2,
\end{align}
as estimated in the main text.

\section*{Complete multipolar picture}

In the diagonalization done in the methods section of the main text, the first canonical transformation modifies the sphere and field momenta, while leaving the emitter's momentum unchanged. This way, we are effectively performing a ``partial'' multipolar coupling transformation, whereby the emitter remains in minimal coupling. Here, we show the result of performing a total transformation~\cite{cohenQEDIntro2007}. Neglecting magnetic effects, the new Hamiltonian is
\begin{align}
    \nonumber
    H &= H_\mathrm{s} + H_\mathrm{s} + H_\mathrm{f} + H_\mathrm{int}\\
    \nonumber
    H_\mathrm{e} &= \sum_{i\in\mathrm{e}} \frac{\mathbf{p}^2_i}{2m_i} + \sum_{i>j\in\mathrm{e}}\frac{q_i q_j}{4\pi\varepsilon_0 |\mathbf{r}_i - \mathbf{r}_j|} + \frac{1}{2\varepsilon_0}\int\mathrm{d}^3r\ \left(\mathbf{P}_\mathrm{e}^\perp(\mathbf{r})\right)^2\\
    \nonumber
    H_\mathrm{s} &= \frac{p_\mathrm{s}^2}{2m_\mathrm{s}} + \frac{m_\mathrm{s}}{2}\Omega_\mathrm{p}^2 z_\mathrm{s}^2\\
    \nonumber
    H_\mathrm{f} &= \int\mathrm{d}^3r\ \left[\frac{\left(\mathbf{D}^\perp(\mathbf{r})\right)^2}{2\varepsilon_0} + \frac{\varepsilon_0 c^2}{2}\left(\nabla\times\mathbf{A}^\perp(\mathbf{r})\right)^2\right]\\
    H_\mathrm{int} &= -\frac{1}{\varepsilon_0} \int\mathrm{d}^3r\ \left(\mathbf{P}_\mathrm{e}^\perp(\mathbf{r}) + \mathbf{P}_\mathrm{s}^\perp(\mathbf{r})\right)\cdot \mathbf{D}^\perp(\mathbf{r}),
\end{align}
where the emitter and sphere polarization fields are
\begin{subequations}
    \begin{align}
        \mathbf{P}_\mathrm{e}(\mathbf{r}) &= \sum_{i\in\mathrm{e}} q_i \mathbf{r}_i \int_0^1\mathrm{d}\sigma\ \delta(\mathbf{r}-\sigma\mathbf{r}_i)\\
        \mathbf{P}_\mathrm{s}(\mathbf{r}) &\simeq -\rho(\mathbf{r})z_\mathrm{s} \hat{\mathbf{z}}.
    \end{align}
\end{subequations}
In \(\mathbf{P}_\mathrm{s}(\mathbf{r})\), we have used that \(\nabla\cdot\left(\dot{\mathbf{P}}_\mathrm{s}(\mathbf{r}) - \mathbf{j}_\mathrm{s}(\mathbf{r})\right)\) to find \(\mathbf{P}_\mathrm{s}^\parallel(\mathbf{r})\), and then chosen its transverse part such that \(\dot{\mathbf{P}}_\mathrm{s}^\perp(\mathbf{r}) = \mathbf{j}_\mathrm{s}^\perp(\mathbf{r})\) as well. This is a valid choice, as the transverse part of the polarization is not a physical quantity and can be chosen freely. It, however, plays a big role in the precise form of the transformed Hamiltonian. The field's new canonical momentum is now essentially the displacement field \(\boldsymbol{\Pi}^\perp(\mathbf{r}) = -\mathbf{D}^\perp(\mathbf{r}) = -\left(\varepsilon_0 \mathbf{E}^\perp(\mathbf{r}) + \mathbf{P}_\mathrm{e}^\perp(\mathbf{r}) + \mathbf{P}_\mathrm{s}^\perp(\mathbf{r})\right)\), and the emitter and sphere Hamiltonians acquire their polarization self-energies. A particularly noteworthy feature of this Hamiltonian is that there is no explicit emitter-sphere direct coupling term: the Coulomb interaction appears to be missing. This is, however, an artifact of the reshuffling of light and matter degrees of freedom involved in the multipolar coupling picture compared to the minimal coupling picture. Indeed, the canonical momentum of the field is not a purely photonic quantity, but it includes the charges' polarization fields as well.

\begin{figure}[htbp]
    \centering
    \includegraphics[scale=0.5]{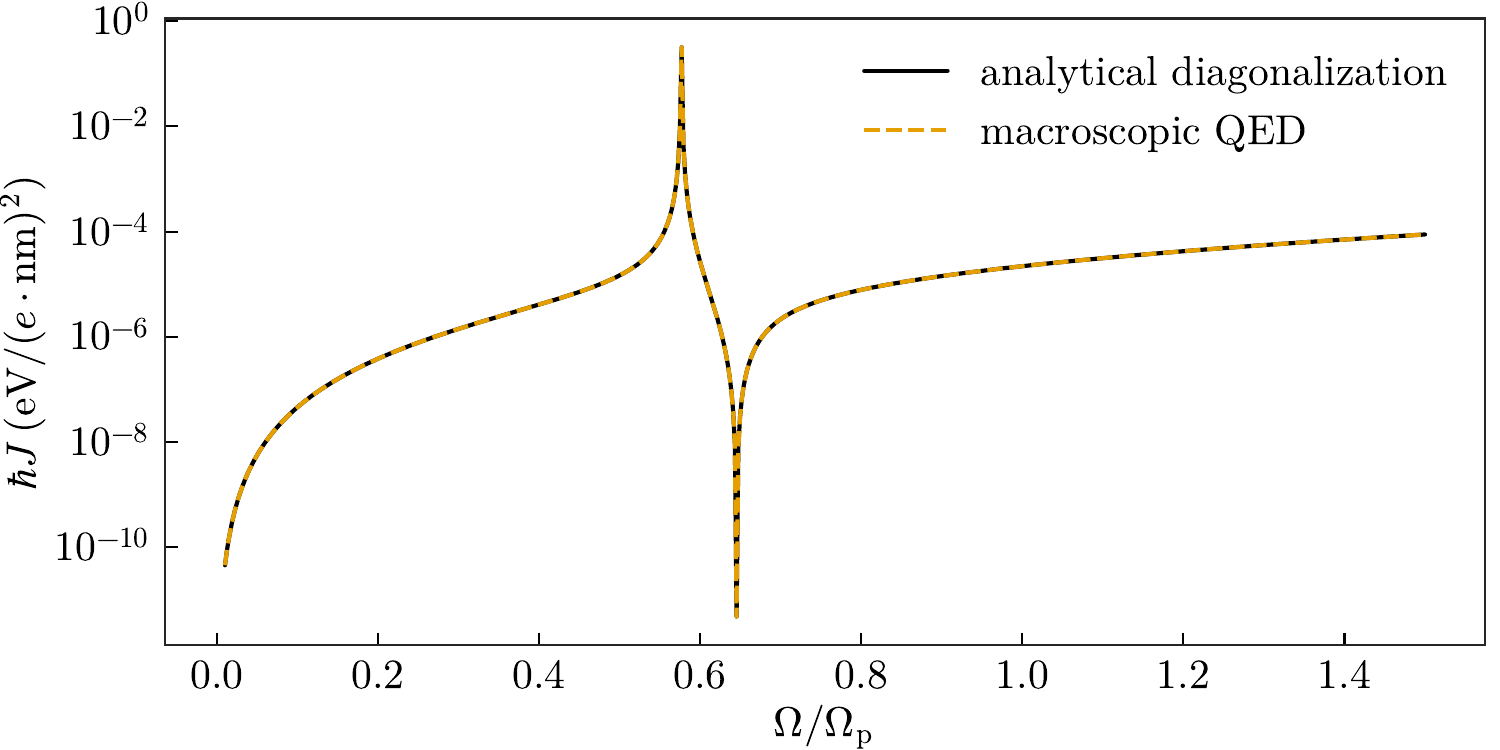}
    \caption{Spectral density per unit dipole moment squared, calculated through the analytical diagonalization and through the macroscopic QED expression. The parameters are \(R_\mathrm{s}=2\)~nm and \(|\mathbf{r}_\mathrm{e}|=4\)~nm.}
    \label{fig.comparison}
\end{figure}

We may now proceed similarly to the main text, by extracting the sphere-photon subsystem and diagonalizing it. It turns out that the mathematical procedure is identical to the one in the main text, the real difference being only the physical interpretation of the dynamical variables. For brevity, we will only show the result for the emitter's total interaction strength, given by part of \(H_\mathrm{int}\) above, in the new polaritonic eigenbasis. The corresponding spectral density will then be compared to its macroscopic QED analog, to provide a supplementary check the validity of our calculations. In the long-wavelength approximation, for an emitter placed along the \(z\) axis and whose dipole moment points along the same direction, we have
\begin{align*}
    H^\mathrm{e}_\mathrm{int} &= -\mathbf{d}\cdot \int\mathrm{d}\Omega\ \left[\frac{1}{\sqrt{\varepsilon_0}}\int\mathrm{d}\omega\ \mathbf{f}_{10}^\mathrm{(tm)}(\omega, \mathbf{r}_\mathrm{e})c_2(\Omega, \omega)\omega\right]\beta(\Omega) = -\hbar\sum_t \left(\sigma_t + \sigma_t^\dagger\right)\int\mathrm{d}\Omega\ \sqrt{J_t(\Omega)}\left(b^\dagger(\Omega) + b(\Omega)\right).
\end{align*}
Here, 
\begin{align}
    \sqrt{J_t(\Omega)} = \frac{|\mathbf{d}_t|}{|\mathbf{r}_\mathrm{e}|} c_1(\Omega) \sqrt{\frac{3 }{2\pi^2\hbar\Omega\varepsilon_0 c}}\int\mathrm{d}\omega\ j_1\left(\frac{\omega |\mathbf{r}_\mathrm{e}|}{c}\right) \bigg[\mathrm{PV}\frac{1}{\Omega^2-\omega^2} + \frac{\Omega^2-\Omega_\mathrm{p}^2-F(\Omega)}{\gamma^2(\Omega)}\delta(\Omega-\omega)\bigg]\gamma(\omega) \omega.
\end{align}
The second term in the square brackets yields
\begin{align*}
    \sqrt{J^{(2)}_t(\Omega)} = \frac{|\mathbf{d}_t|}{|\mathbf{r}_\mathrm{e}|} \sqrt{\frac{3\Omega }{2\pi^2\hbar\varepsilon_0 c}} j_1\left(\frac{\Omega |\mathbf{r}_\mathrm{e}|}{c}\right) \frac{\Omega^2-\Omega_\mathrm{p}^2-F(\Omega)}{\gamma(\Omega)} c_1(\Omega),
\end{align*}
while the first one is
\begin{align*}
    \sqrt{J_t^{(1)}(\Omega)} 
    &= 2\Omega_\mathrm{p} \frac{|\mathbf{d}_t|}{|\mathbf{r}_\mathrm{e}|}c_1(\Omega) \sqrt{\frac{3 R_\mathrm{s}}{2\pi^3\hbar\Omega\varepsilon_0 c^2}}\mathrm{PV}\int\mathrm{d}\omega\ \frac{\omega^2j_1\left(\frac{\omega R_\mathrm{s}}{c}\right)j_1\left(\frac{\omega |\mathbf{r}_\mathrm{e}|}{c}\right)}{\Omega^2-\omega^2} \\
    &= 
    -\Omega_\mathrm{p}\frac{|\mathbf{d}_t|}{|\mathbf{r}_\mathrm{e}|} c_1(\Omega) \sqrt{\frac{3 }{\pi^2\hbar\Omega\varepsilon_0 R_\mathrm{s}}}\mathcal{F}\left[
        j_1(x) j_1\left(\frac{|\mathbf{r}_\mathrm{e}|}{R_\mathrm{s}}x\right) +
        \left(\frac{\Omega R_\mathrm{s}}{c}\right)^2\frac{j_1(x) j_1\left(\frac{|\mathbf{r}_\mathrm{e}|}{R_\mathrm{s}}x\right)}{x^2 - \left(\frac{\Omega R_\mathrm{s}}{c}\right)^2}
    \right]\\
    &= \frac{|\mathbf{d}_t|}{|\mathbf{r}_\mathrm{e}|} \sqrt{\frac{3 \Omega R_\mathrm{s}}{2\pi^2\hbar\varepsilon_0 c}} \Omega_\mathrm{p} \sqrt{\frac{\pi}{c}} j_1\left(\frac{\Omega R_\mathrm{s}}{c}\right)y_1\left(\frac{\Omega |\mathbf{r}_\mathrm{e}|}{c}\right)c_1(\Omega).
\end{align*}
Added together, both terms result in the full spectral density:
\begin{equation}
    J_t(\Omega) = \left(\sqrt{J^{(1)}_{t}(\Omega)} + \sqrt{J^{(2)}_{t}(\Omega)}\right)^2,
\end{equation}
which physically contains both emitter-sphere Coulomb interactions and emitter-photon transverse interactions. We may plot in \autoref{fig.comparison} the above expression and compare it to its macroscopic QED counterpart, given by
\begin{equation}
    J_t^\mathrm{MQED}(\Omega) = \frac{\Omega^2 }{\hbar\pi\varepsilon_0 c^2} \hat{\mathbf{z}}\cdot \mathrm{Im}\mathbf{G}(\mathbf{r}_\mathrm{e}, \mathbf{r}_\mathrm{e}, \Omega)\cdot \hat{\mathbf{z}},
\end{equation}
where \(\mathbf{G}\) is the EM Green tensor of the sphere.
For the comparison, we have calculated the Green tensor for a metallic sphere with a lossless Drude permittivity compatible with the parameters used in the main text (\(\rho=58.9 e/\mathrm{nm}^3\)), and we have restricted the Green tensor evaluation to only include the \(l=1\) order. The agreement between both methods is extraordinary, which validates the manipulations and approximations made throughout the article and this supplementary information.

\bibliography{references}